\documentclass[iop]{emulateapj}
\usepackage{amsmath,color}
\usepackage{url}

\newcommand{\ltaraw}{$\; \buildrel < \over \sim \;$}
\newcommand{\lta}{\lower.5ex\hbox{\ltaraw}}
\newcommand{\gtaraw}{$\; \buildrel > \over \sim \;$}
\newcommand{\gta}{\lower.5ex\hbox{\gtaraw}}

\newcommand{\beq}{\begin{equation}}
\newcommand{\eeq}{\end{equation}}
\newcommand{\beqa}{\begin{eqnarray}}
\newcommand{\eeqa}{\end{eqnarray}}
\newcommand{\om}{\Omega_m}

\shorttitle{Robust Redshift Space Distortions}
\shortauthors{Kwan, Lewis, Linder}

\begin{document}


\title{Mapping Growth and Gravity with Robust Redshift Space Distortions\footnotemark[1]}

\author{Juliana Kwan}
\affil{High Energy Physics Division, Argonne National Laboratory, Lemont, IL, 60439}
\affil{Sydney Institute for Astronomy, School of Physics, A28, The University of Sydney, NSW 2006, Australia}

\author{Geraint F. Lewis}
\affil{Sydney Institute for Astronomy, School of Physics, A28, The University of Sydney, NSW 2006, Australia}

\and

\author{Eric V.\ Linder}
\affil{Lawrence Berkeley National Laboratory \& University of California, Berkeley, CA 94720, USA}
\affil{Institute for the Early Universe WCU, Ewha Womans University, Seoul, Korea}



\begin{abstract}
Redshift space distortions caused by galaxy peculiar velocities
provide a window onto the growth rate of large scale structure and a
method for testing general relativity.  We investigate through a
comparison of N-body simulations to various extensions of perturbation
theory beyond the linear regime, the robustness of cosmological
parameter extraction, including the gravitational growth index
$\gamma$.  We find that the Kaiser formula and some perturbation
theory approaches bias the growth rate by $1\sigma$ or more relative
to the fiducial at scales as large as $k > 0.07\,h$/Mpc. This bias
propagates to estimates of the gravitational growth index as well as
$\Omega_m$ and the equation of state parameter and presents a
significant challenge to modelling redshift space distortions. We also
determine an accurate fitting function for a combination of line of
sight damping and higher order angular dependence that allows robust
modelling of the redshift space power spectrum to substantially higher
$k$.
\end{abstract}

\section{Introduction}     \label{intro} 
Large scale structure surveys provide key pieces of evidence for the
accelerated expansion of the universe and historically have made
substantial contributions to establish $\Lambda$-Cold Dark Matter
($\Lambda$CDM) as the current standard model of
cosmology. Measurements of the baryon acoustic oscillation (BAO) scale
and the shape of the matter power spectrum continue to be essential
probes of cosmology.  But although the amount of cosmological data
grows, the presence of a cosmological constant $\Lambda$, or other
forms of dark energy, has yet to be confirmed by observations over
much of cosmic time (from $1 < z < 1100$) and some highly undesirable
features of $\Lambda$CDM, such as the cosmic coincidence problem and a
discrepancy of 120 orders of magnitude in the value of
$\Lambda$~\citep{2001LRR.....4....1C}, are still unsolved.

\let\thefootnote\relax\footnotetext{\footnotemark[1]Research undertaken as
  part of the Commonwealth Cosmology Initiative (CCI: www.thecci.org),
  an international collaboration supported by the Australian Research
  Council.}

A possible alternative explanation of cosmic acceleration suggests
that the nature of gravity deviates from general relativity on large
scales, thus affecting the measurement of the distance-redshift
relation and the manner in which large scale structure
forms. Recently, the redshift space distortions (RSD) seen in large
scale structure surveys have emerged as a powerful new method of
probing such a gravitational origin~\citep{2008APh....29..336L}
because they offer a means of measuring the growth rate of large scale
structure directly.

Currently, the most developed probes of cosmic acceleration originate
from distance measurements but say very little about gravitational
dynamics. However, a constraint on the growth rate may assist in
breaking a possible degeneracy between models with the same expansion
history but differing in terms of their gravitational physics. Few
probes are as sensitive to gravitational dynamics as RSD; weak lensing
measures the integrated growth to some redshift, but not the growth
rate.  Pairwise velocity statistics of objects, e.g.\ supernovae, have
been proposed to explore the growth rate, but this technique will not
produce competitive constraints until a sufficiently high number
density of objects are measured with a future survey such as the Large
Synoptic Survey Telescope~\citep{1008.2560}.

The growth rate can be measured via redshift space distortions to
modest precision in recent and ongoing large galaxy redshift surveys,
such as SDSS-II \citep{2009ApJS..182..543A}, VVDS
\citep{2005A&A...439..845L}, WiggleZ \citep{2010MNRAS.401.1429D}, and
BOSS \citep{1101.1529} that cover $z=0.1$--0.9.  In the next
generation, surveys such as BigBOSS \citep{0904.0468},
SuMIRe~\citep{1007.1256} and Euclid \citep{0912.0914} will provide
precise measurements over an even greater range of redshifts up to
$z=2$.

In Sec.~\ref{sec:vel} we review the physics of redshift space
distortions from the galaxy velocity field.  Section~\ref{sec:rsd}
discusses methods for accurately extracting the growth rate from the
galaxy power spectrum.  To compare analytic approximations from
perturbation theory to the fully non-linear solutions, we carry out
N-body simulations in Sec.~\ref{nbody}.  We then investigate in
Sections~\ref{sec:kaiser},~\ref{sec:quasi} and ~\ref{sec:morept}, the
bias induced in the growth rate, due to the effects of non-linearity
in the density field, as a function of the maximum wavenumber
$k_{max}$. These investigations are split into three classes of
models; the Kaiser limit in Sec.~\ref{sec:kaiser}, the quasi-linear
Scoccimarro ansatz in Sec.~\ref{sec:quasi} and the non-linear models
given by perturbation theory in Sec.~\ref{sec:morept}. We perform a
closer investigation of the appropriate form of the damping function
and angular dependence in the redshift space power spectrum in
Sec.~\ref{sec:damp}, finding an accurate, scale dependent correction
factor, and in Sec.~\ref{sec:extended} we open the parameter space to
discuss the implication of systematics on detecting deviations from
general relativity. The conclusions of this paper are presented in
Sec.~\ref{sec:concl}.

\section{Galaxy Velocity Fields} \label{sec:vel} 

On large scales, the peculiar velocities of galaxies are dominated by
the bulk flow motions induced by the gradients of gravitational potentials. 
These in turn arise from mass density fluctuations and so the velocity 
field traces the growth rate of large scale
structure. The continuity equation gives the relationship
between the density and velocity fields and in the linear density regime is
written as:
\begin{equation}
\nabla \cdot v({\bf x}) = -f\,\delta({\bf x}),
\label{continuity}
\end{equation}
where the $\delta$, the density perturbation is defined in terms of
the density $\rho$ and its mean value $\bar{\rho}$ such that $\delta =
(\rho-\bar{\rho})/\bar{\rho} \ll 1$, $D$ is the linear growth factor
so $\delta(a)\propto D(a)$, and $f \equiv d\ln D/d\ln a$ is the linear
growth rate.

The peculiar velocities add an extra component to the cosmological
redshift which perturbs the real space positions of galaxies, $x_r$,
along the line of sight:
\begin{equation}
x_s = x_r + (1+z)\frac{{\bf v_{pec}\cdot \hat{x}}}{H(z)}
\label{conversion}
\end{equation}
where $x_s$ is the redshift space position and $\hat{x}$ is the
direction of the line of sight.  The expansion rate, or Hubble
parameter, $H(a)=d\ln a/dt$ and the redshift $z=a^{-1}-1$.  This
apparent change in position produces an additional anisotropic
component to the power spectrum or correlation function because the
amount of shifting that occurs is dependent on the angle made with
respect to the line of sight.

The growth rate $f$ can be extracted from measurements of the mass
density power spectrum, in the combination $f(z)\,\sigma_8(z)$
\citep{2009MNRAS.393..297P}, where $\sigma_8$ is the rms mass
fluctuation amplitude proportional to the growth factor $D(a)$.  Thus,
$f\sigma_8\propto dD/d\ln a$.  Within general relativity, the growth
is determined by the expansion history $H(a)$ (assuming negligible
dark energy clustering or interaction).  To separate the effects of
the expansion from any modifications to the standard gravity picture,
\citet{2005PhRvD..72d3529L} introduced the gravitational growth index
$\gamma$,   which parameterises the true linear growth rate $f$ as
\begin{equation}
f \approx \Omega_m(a)^\gamma \,,
\label{def}
\end{equation} 
for models that are matter dominated at high redshift and has been
shown to be accurate to the subpercent level (even better for
$f\sigma_8$) for a number of classes of cosmological models.

The gravitational growth index $\gamma$ provides a useful extension to
the cosmological model framework, allowing straightforward tests of
growth vs expansion, and a compact method for distinguishing between
many classes of modified gravity models.  For instance, a $\Lambda$CDM
model has $\gamma$=0.55, with almost no dependence on the dark energy
equation of state $w$, whereas for a Dvali Gabadadze Porrati (DGP)
gravity model~\citep{2000PhLB..485..208D}, $\gamma$ = 0.68 and for many
$f(R)$ gravity models $\gamma\approx 0.42$ today.  We will explore
therefore not only the measurement of the growth rate $f$, but also
its propagation into cosmological parameter estimation of the matter
density $\om$, the dark energy equation of state, and the
gravitational growth index $\gamma$.

Current measurements of $f$ have been made from galaxy redshift
surveys from the 2dF ~\citep{2003MNRAS.346...78H} and SDSS
[\cite{arXiv:1102.1014} and older measurements] at low redshift; and
at higher redshifts using VVDS~\citep{2008Natur.451..541G} and
WiggleZ~\citep{1104.2948}. These have constrained $f\sigma_8$ at
particular redshifts to $\sim 10$\% at best, but the precision and
redshift coverage are not yet sufficient to stringently test for
modifications to gravity.  Nonetheless, they are exciting precursors
to the constraints that will be produced by BOSS and BigBOSS.  As the
measurements get more precise, it is important to ensure that the
analysis methods and theoretical knowledge keep pace in their
accuracy.

\cite{2008APh....29..336L} demonstrated that once $f$ was extracted 
from next generation measurements of redshift space distortions one 
could place significant constraints on $\gamma$ and theories of gravity. 
Propagating errors, one sees that at a single redshift the relation 
between the uncertainties on $f$ and $\gamma$ is 
\beq 
\frac{\delta f}{f}=\gamma\, \frac{\delta\om(a)}{\om(a)} + \ln\om(a)\,\, 
\delta\gamma \,. \label{eq:ferror} 
\eeq For an experiment to determine $\gamma$ say to 0.04, we would require
a 2\% measurement of $f$ at $z=0.5$ or a 1\% measurement at $z=1$ if
we knew the expansion history to precisely follow a $\Lambda$CDM
concordance cosmology. Without perfect knowledge of $\Omega_m(a)$, the
constraint on $f$ would need to be tighter to achieve the same
precison on $\gamma$. Thus the ability to measure accurately $f$ from
a survey is a crucial topic to investigate.  This article investigates
the impact of various systematics on the quality of cosmological
constraints as we approach the precision required to produce strong
tests of our understanding of gravity.

Methods for realistically extracting the growth rate from the measured
galaxy power spectrum have been studied by, e.g.,
\cite{2011ApJ...726....5O} in the linear regime and
\cite{2009MNRAS.393..297P, 2011MNRAS.410.2081J} in the quasi-linear
regime.  We extend their analyses by investigating more fully the
effects of including non-linearities and their impact on a broader set
of cosmological parameters that are expected to influence measurements
of the growth rate. We test how well various models for redshift space
distortions perform, focusing on three redshifts, $z=0,0.5,1$ with
three cuts in scale; $k_{max} = 0.07\,h$/Mpc to test RSD on large
scales well in the linear regime, $k_{max}=0.1\,h$/Mpc, at the onset
of non-linearity and $k_{max}=0.2\,h$/Mpc, to investigate the
non-linear regime, although it is really distinctions such as these
that we are aiming to probe.

\section{Redshift Space Distortion Theory} \label{sec:rsd} 

Redshift space distortions introduce an anisotropic component to the
power spectrum, as the peculiar velocity of the galaxy projected along
the line of sight adds to the cosmological redshift, perturbing the
galaxy positions. This occurs as a bulk effect and to be able to
reliably extract the growth rate from it we need to measure a
statistic such as the power spectrum of density fluctuations or the
spatial correlation function, which quantify the degree to which
objects cluster.  In this work, we consider the power spectrum because
of its close relationship with theory and the ease with which a linear
power spectrum may be obtained from a Boltzmann code such as
CAMB~\citep{2000ApJ...538..473L}.  Other works have utilised the
correlation function instead to investigate RSD, such
as~\cite{2008Natur.451..541G, 1101.2608}.

\begin{deluxetable*}{lclcl}
\tabletypesize{\scriptsize} \tablewidth{0pt} \tablecaption{List of
models considered, their free parameters and mentions in other
literature. We have divided this table into three sections,
corresponding to Sections~\ref{sec:kaiser}, \ref{sec:quasi}
and~\ref{sec:morept} as we progress down the table. Note that
empirical damping means that we allow the MCMC process to decide the
amount of damping necessary for a good fit.\label{modelnames}}

\tablehead{
\colhead{Model}           &
\colhead{Parameters}      &   
\colhead{Equation}        & 
\colhead{}                &
\colhead{Reference} }
\startdata
Kaiser       & $f, b$  & $P^s(k,\mu) = \left(b+f\mu^2\right)^2 P_L(k)$                       & [\ref{Kaiser}]     & \cite{1987MNRAS.227....1K}\\
Streaming    & $f, b$  & $P^s(k,\mu) = e^{-(fk\mu\sigma_v)^2}\left(b+f\mu^2\right)^2 P_L(k)$ &[\ref{streaming}]   & \cite{1995ApJ...448..494F}\\
Empirical    & $f, b,\sigma_v$   & As above but $\sigma_v$ is a free parameter                 &[\ref{streaming}] & \cite{arXiv:1102.1014}   \\
Non-linear $P_{\delta\delta}$ & $f,b$   & As for streaming but $P_{L}$ is replaced by non-linear $P_{\delta\delta}$&[\ref{streaming}]  & \cite{1104.2948} \\
\tableline
Scoccimarro  & $f, b$         & $P^s_{q}(k,\mu) = e^{-(fk\mu\sigma_v)^2} \times$               & [\ref{scoccimarro}] & \cite{2004PhRvD..70h3007S}\\
 (with linear damping)  &     & $\qquad \left[b^2 P_{\delta\delta} (k) - 2 \mu^2 P_{\delta\theta} + \mu^4 P_{\theta\theta}\right] $ &    & \\
Scoccimarro          & $f, b, \sigma_v$   & As above but $\sigma_v$ is a free parameter & [\ref{scoccimarro}]      & \cite{2004PhRvD..70h3007S}\\
  (with empirical damping)  &        &                                           &            & \cite{2011ApJ...727L...9J}\\
\tableline
SPT          & $f, b$  & $P^s_{SPT}(k,\mu) = (b+f\mu^2)^2 P_L + (b+f\mu^2)P^s_{13} + P^s_{22}$ &[\ref{SPT}]        & \cite{1998MNRAS.301..797H}\\
      &  &                                                                &            & \cite{1999ApJ...517..531S}\\
      &  &                                                                &            & \cite{2008PhRvD..77f3530M}\\
LPT          & $f, b$  & $P^s_{LPT} = e^{-k^2(1+f(f+2)\mu^2)\sigma_v^2} \times $               &  [\ref{LPT}]      & \cite{2008PhRvD..77f3530M}\\
      &  & $\qquad \left[P^s_{SPT} +  (b+f\mu^2)^2 P_L k^2(1+f(f+2)\mu^2)\sigma_v^2\right]$ & &                    \\
Taruya$^{++}$& $f, b$  & $P^s(k,\mu) = P^s_{q} + e^{-(fk\mu\sigma_v)^2}\times $                & [\ref{closure}]   & \cite{2010PhRvD..82f3522T}\\
(with linear damping) &  & $\qquad \left[b^3 A(k,\mu, f, b) + b^4 B(k,\mu,f,b) \right]$ &            &                    \\
Taruya$^{++}$& $f, b, \sigma_v$ & As above but $\sigma_v$ is a free parameter                  & [\ref{closure}]   & \cite{2010PhRvD..82f3522T}\\
(with empirical damping) &  &                                                    &            &                    \\
\enddata
\end{deluxetable*}

The complete, non-linear redshift space power spectrum is given by:
\begin{equation}
P^s({\bf k}) = \int \frac{d^3 r}{(2\pi)^3} e^{-i{\bf k \cdot r}} \left< e^{ifk\mu\Delta u_z} \left[ 1 + \delta({\bf x})\right]\left[1 + \delta({\bf x'})\right]\right>,
\label{fullRSD}
\end{equation} 
as presented in~\cite{2004PhRvD..70h3007S} where ${\bf k}$ is the
wavevector, ${\bf r} = {\bf x'}- {\bf x}$, $\Delta u_z = u({\bf x'}) -
u({\bf x})$ and $u$ is the peculiar velocity field expressed in
comoving coordinates such that $u = v_{pec}/H$. Even though we can
write down the full expression, it involves quantities, such as the
non-linear density contrast $\delta({\bf{x}})$, that we cannot easily
link to theoretical predictions when given a set of cosmological
parameters, and thus the growth rate cannot be extracted this way.

In the linear density limit, the above equation reduces to something tractable
known as the Kaiser formula~\citep{1987MNRAS.227....1K}:
\begin{equation}
P^s(k,\mu) = \left(b+f\mu^2\right)^2 P_{\delta\delta}(k), 
\label{Kaiser}
\end{equation}
where $\mu$ is the cosine of the angle made by ${\bf k}$ with respect
to the line of sight (${\bf r}$ in Eq.~\ref{fullRSD}), thus making an
essentially two-dimensional power spectrum involving radial (line of
sight) and transverse modes.  This anisotropy is in contrast to the
undistorted linear matter power spectrum $P_{\delta\delta}$.  Note $b$
is the linear bias factor, relating the galaxy density fluctuations to
the mass fluctuations.  Equation~\ref{Kaiser} is a result of several
assumptions imposed on the full relationship between real and redshift
space in Eq.~\ref{fullRSD}, which we explore in Sec.~\ref{sec:kaiser}.

To extend redshift space distortion theory to smaller scales, several
quasi-linear and non-linear models have been proposed. The full linear
model for redshift space distortions includes a velocity streaming term 
in the form of an extra exponential damping term, 
\begin{equation}
P^s(k,\mu) = e^{-(fk\mu\sigma_v)^2}(b+f\mu^2)^2 P_{\delta\delta},
\label{streaming}
\end{equation}
where $\sigma_v^2 = [1/(6\pi^2)]\int \, P_L \, dk$ and $P_L$ is the
linear power spectrum~\citep{1995ApJ...448..494F}.  Within the
Kaiser limit, this multiplication/convolution accounts for the joint
density and velocity probability distributions along the line of sight
in Fourier/real space assuming that the pairwise velocity probability
distribution function (PDF) is Gaussian.

\cite{1995ApJ...448..494F} inspired others to take a similar approach
to modelling the more non-linear Fingers-of-God (FoG)
effect~\cite[see][for early
discussions]{1972MNRAS.156P...1J,1977ApJ...212L...3S}, and an
additional small scale velocity dispersion term $\sigma_{vir}$ arises
from convolving the pairwise velocity dispersion profile of objects
within a halo with the linear Kaiser or streaming model. Assuming a
Gaussian profile for random motion of galaxies within virialised
structures results in the exponential damping term proposed
by~\cite{1994MNRAS.267.1020P}, while using an exponential profile for
the velocity dispersion, first derived by~\cite{1976Ap&SS..45....3P}
and then applied to the power spectrum by~\cite{1994ApJ...431...569P},
results in a Lorentzian damping term. Such a combination is often then
taken to be the full quasi-linear model of redshift space
distortions~\citep{2010PhRvD..81b3526D}
\begin{equation}
P^s(k,\mu) = e^{-(fk\mu\sigma_v)^2}V_{vir}(k,\mu)(b+f\mu^2)^2 P_{\delta\delta} 
\,, 
\label{quasilinear}
\end{equation}
where $V_{vir} = \exp[-(k\sigma_{vir}\mu)^2]$ for the Gaussian profile
or $V_{vir} = \left(1+k^2\sigma_{vir}^2\mu^2\right)^{-1}$ for
Lorentzian damping. These are of course the same to first order. Note
that $\sigma_v$ and $\sigma_{vir}$ are not identical in general
because $\sigma_v$ quantifies the bulk motion of objects, for example
haloes, while $\sigma_{vir}$ aims to model small scale motion, such as
that of bound objects within a halo.

However, Eqs.~\ref{streaming} and~\ref{quasilinear} are still linear
in the sense that the densities and velocities are assumed to be
exactly coherent such that the bulk flow of the dark matter traces
out the velocity divergences exactly and $P_{\theta\theta} =
f^2P_{\delta\delta}$ and $P_{\delta\theta} = -fbP_{\delta\delta}$. The
simplest of the models that accounts for the additional information
contained in the non-linear power spectrum of velocity divergences is
the ansatz proposed in \cite{2004PhRvD..70h3007S}, 
\begin{equation}
P^s(k,\mu) = e^{-(fk\mu\sigma_v)^2}\left[b^2 P_{\delta\delta} (k) - 2 \mu^2 P_{\delta\theta} + \mu^4 P_{\theta\theta}\right], 
\label{scoccimarro}
\end{equation}
where $P_{\delta\delta}$, $P_{\delta\theta}$ and $P_{\theta\theta}$
are the {\em non-linear} density, density-velocity and velocity
divergence power spectra, defined as $P_{\delta\theta} =
\left<|\,\delta\theta^*|\right>$ and $P_{\theta\theta} = \left<\,|\theta|^2\right>$, where
$\theta = \nabla \cdot {\bf v}$. The velocity dispersion is given by
$\sigma_v^2$ as defined in
Eqn.~\ref{streaming}. \cite{2011ApJ...727L...9J} demonstrated that
this recovers the linear growth rate up to $k = 0.25\,h$/Mpc to a
precision on $f$ of 0.64\%, but it is difficult to predict
$P_{\delta\delta}, P_{\delta\theta}$ and $P_{\theta\theta}$ in the
fully non-linear regime without recourse to N-body simulations.

One of the most promising avenues for treating redshift space
distortions is to take a perturbative approach to the full
transformation (Eq.~\ref{fullRSD}) and expand the term within angular
brackets.  The most widely used and simplest perturbative scheme is
Standard Perturbation Theory (SPT), in which the expansion is
performed in powers of the scale factor~\citep[see][for more
details]{1994ApJ...431..495J}.  Such schemes are expected to model the
non-linear effects to higher accuracy than Eq.~\ref{scoccimarro},
since that model does not constitute an exact expression under
perturbation theory as previously noted
by~\cite{2008PhRvD..77f3530M,2010PhRvD..82f3522T}. Some caution is
required; no single perturbative scheme is preferred, they each have
their regimes of validity~\citep{2009PhRvD..80d3531C}, and they are
limited to being useful only when the situation is mostly linear but
requires a small correction. The demarcation between linear and
non-linear regimes is often different in redshift and real space and
some perturbation theories have a more limited range of accuracy in
redshift space~\citep{1999ApJ...517..531S}. Some of these problems are
addressed by using other perturbative schemes such as Renormalised
Perturbation Theory (RPT) and closure theory.

Perturbation theory adds additional terms to the Kaiser formula to
model the non-linear contributions to the redshift space power
spectrum. The next to linear order in the SPT expansion of the real
space power spectrum is the 1-loop expression given by:
\begin{equation}
P(k) = P_{L} + P_{13} + P_{22}, 
\end{equation}
where $P_{L}$ is the linear power spectrum, and $P_{13}, P_{22}$ are
mode coupling terms containing extra corrections to third order in
$\delta_k$. For redshift space, these terms are:
\begin{equation}
P^s(k,\mu) = (b+f\mu^2)^2 P_L + (b+f\mu^2)P^s_{13} + P^s_{22}
\label{SPT}
\end{equation}
as derived by \cite{1998MNRAS.301..797H, 1999ApJ...517..531S,
  2008PhRvD..77f3530M} and we refer the reader to these works for the
  complete expression. Both mode coupling terms in Eqn.~\ref{SPT} now
  contain additional redshift space contributions that are sensitive
  to the growth rate. In Lagrangian Perturbation Theory (LPT), this
  becomes
\begin{equation}
\begin{split}
P^s(k,\mu) = e^{-k^2(1+f(f+2)\mu^2)\sigma_v^2}\left[(b+f\mu^2)^2 P_L + (b+f\mu^2)P^s_{13} \right.\\
\left. + P^s_{22} + (b+f\mu^2)^2 P_L k^2(1+f(f+2)\mu^2)\sigma_v^2\right],
\label{LPT}
\end{split}
\end{equation}
as derived by~\cite{2008PhRvD..77f3530M}. An exponential damping term
naturally arises in the derivation that accounts for both BAO damping
and RSD smearing.  Although some of the mode coupling terms look
formidable, these equations are not too difficult to derive and
evaluate. However, the expressions beyond the 1-loop terms are much
more computationally demanding; the presence of mode coupling produces
multidimensional integrals that need to be performed numerically and
that can be as time consuming as running an entire N-body
simulation. This is far too expensive to be done in the course of a
parameter fitting routine during which the model power spectrum may
need to be evaluated at each point in parameter space.

Another popular perturbation scheme is closure theory, in which higher
order N-point statistics, such as the bispectrum, are defined in terms
of derivatives of lower order N-point statistics to close the system
of equations. The closure theory expansion of the redshift space power
spectrum derived by~\cite{2010PhRvD..82f3522T}, (hereafter known as
Taruya$^{++}$), produces additional terms in the Scoccimarro formula,
and Eq.~\ref{scoccimarro} is now
\begin{equation}
\begin{split}
P^s(k,\mu) = e^{-(fk\mu\sigma_v)^2}\left[b^2 P_{\delta\delta} (k) - 2 \mu^2 P_{\delta\theta} + \mu^4 P_{\theta\theta}+ \right.\\
\left. b^3 A(k,\mu, f, b) + b^4 B(k,\mu,f,b) \right], 
\label{closure}
\end{split}
\end{equation}
where $A(k,\mu,f,b)$ and $B(k,\mu,f,b)$ contain the extra
contributions to the power spectrum arising from the coupling of the
density and velocity fields (see \cite{2010PhRvD..82f3522T} for the
full terms), which are rather lengthy to calculate even in an era
where computational power is cheap. These three models, SPT, LPT and
closure theory, comprise the main perturbative approaches to modelling
the RSD power spectrum.

\section{N-body simulations}\label{nbody}

Finding a robust approach to extracting growth rate information
requires comparison to the non-linear power spectra as given by
simulations.  The simulations were performed using
GADGET2~\citep{2005MNRAS.364.1105S} with a box of length 1500 Mpc/$h$
per side and 1024$^3$ particles and a particle mass of $\approx 2.2
\times 10^{11} M_{\odot}/h$. We chose a reference $\Lambda$CDM
cosmology with $\Omega_{m,0} = 0.25$, $\Omega_{\Lambda,0} = 0.75$,
$\Omega_{b,0}=0.045$, $n_s = 0.97$, $\sigma_8 = 0.8$, $h=0.72$.

There are numerous factors that affect the ability of a N-body
simulation to accurately reproduce the growth of structure but we will
only outline the most significant effects that could impact our
results. The box size was chosen to be large enough such that the
growth of the modes on the scales of interest could be fed with a
sufficient number of larger modes via mode coupling. The initial
conditions for the simulations were obtained using the Zel'dovich
approximation~\citep{1970A&A.....5...84Z} evaluated at $z=199$; this
is a sufficiently high redshift to prevent an artificial damping of
the power spectrum that occurs when the modes are not given enough
time for non-linear growth~\citep{2010ApJ...715..104H}. We ran 20
simulations with the same cosmological parameters but each simulation
is initialized by a different realization of a Gaussian field set by
the fiducial cosmology.  We also ran a number of simulations with
different box sizes, particle numbers and error tolerances in the
gravity calculations to ensure numerical convergence. Each simulation
contributes three projections of a redshift distorted density field
from the three lines of sight oriented along each axis of the box,
which we average over as
in~\cite{2008Natur.451..541G,2009MNRAS.393..297P,
2011MNRAS.410.2081J}. Although the density modes of each projection
shared the same growth history, the distribution of particles seen
along each axis is now different and each redshift space power
spectrum contains new information. This gives a final space power
spectrum that has been averaged over 20 realisations with each
simulation contributing three projections of the redshift space power
spectrum. As an accuracy test of our simulations, we have compared our
real space power spectrum averaged over all the realisations to the
power spectrum emulator in~\cite{2010ApJ...713.1322L} and found a
dispersion of less than 2\% between the power spectra (1\% with
Halofit~\citep{2003MNRAS.341.1311S}) over the scales we consider in
this article. The real space density and redshift space power spectra
are measured by applying the method outlined in
\cite{2005ApJ...620..559J} with a third order mass assignment scheme
(triangular shaped cloud, TSC) to smooth the particles on to a FFT
grid size of $2048^3$ for the density power spectra. The
density-velocity and velocity power spectra, $P_{\theta\delta}$ and
$P_{\theta\theta}$, also use a TSC mass assignment scheme but with an
additional step that involves dividing by the densities to remove the
mass weighting imposed by binning the
particles~\citep{2004PhRvD..70h3007S}. Because of the sparseness of
the particle density in certain regions of the simulation, we could
only achieve a maximum FFT grid resolution of $512^3$ for these power
spectra.

The cosmological parameters were derived using the Metropolis
algorithm of CosmoMC~\citep{2002PhRvD..66j3511L} to facilitate a
Markov Chain Monte Carlo (MCMC) process to calculate the likelihood
distribution of these parameters assuming a particular model of RSD is
true. The errors on the redshift space power spectrum were estimated
using $\sim 1200$ Gaussian realisations (depending on how many were
required for the covariance matrix to converge) using the same volume
but only $256^{3}$ particles for speed. Applying the Zel'dovich
approximation on the Gaussian density field then gives the velocities
as:
\begin{equation}
v(x) = aHfD \phi (x),
\label{eqn:zeldovich}
\end{equation}
where ${\bf \phi}$ is the potential for a particle located at
$x$. For the smallest range in $k$, we were able to obtain the
covariance matrix directly from the simulations and this provided a
useful check against the errors given by the Zel'dovich
approximation. We found that the resultant best fitting values for the
growth rate were almost indistinguishable for $0.03 < k<
0.07\,h$/Mpc. A linear theory power spectrum $P_L$ is produced by CAMB
using the Parameterised Post Friedmann (PPF)
module~\citep{2008PhRvD..78h7303F} to treat the perturbations in the
fluid equations for $\{w_0,w_a\}$ models, where the dark energy
equation of state is given by $w(a)=w_0+w_a(1-a)$.  Utilizing the
linear theory power spectrum, we have implemented SPT, LPT and
Taruya$^{++}$ theory schemes for the real and redshift space power
spectra within CosmoMC in order to explore the full parameter range
that affects growth. The growth rate is modelled either in terms of
$f$ or $\gamma$ as an additional free parameter. We consider a chain
to have converged when $R - 1 \leq 0.01 $, where $R$ is defined as the
ratio of the variance between the mean of the chains and mean of the
chain variance.

\section{Limits of the Kaiser Formula} \label{sec:kaiser} 

The Kaiser formula has already been demonstrated to have a limited
range of applicability by ~\cite{2011ApJ...727L...9J,
2011MNRAS.410.2081J, 2011ApJ...726....5O} and most notably by
\cite{2004PhRvD..70h3007S} but despite these shortcomings has been
applied to a number of data sets, most of these at low redshift, deep
in the non-linear regime. \cite{2011ApJ...726....5O,
2011ApJ...727L...9J} showed that the Kaiser formula is unable to
reproduce the growth rate measured from a N-body simulation on all but
the very largest scales, especially for low mass haloes. We revisit
the problem using the original Kaiser formula as presented in
\cite{1987MNRAS.227....1K} but then extend our analysis to address its
various extensions such as the streaming
model~\citep{1995ApJ...448..494F}, and the various quasi-linear FoG
damping models that have been proposed~\citep{1994MNRAS.267.1020P,
1994ApJ...431...569P}. The analysis of the Kaiser limit in this
section provides a point of comparison for the results of the higher
order models that we analyse in Sections~\ref{sec:quasi}
and~\ref{sec:morept}.

\begin{figure}
\begin{center}
\includegraphics[width = \columnwidth]{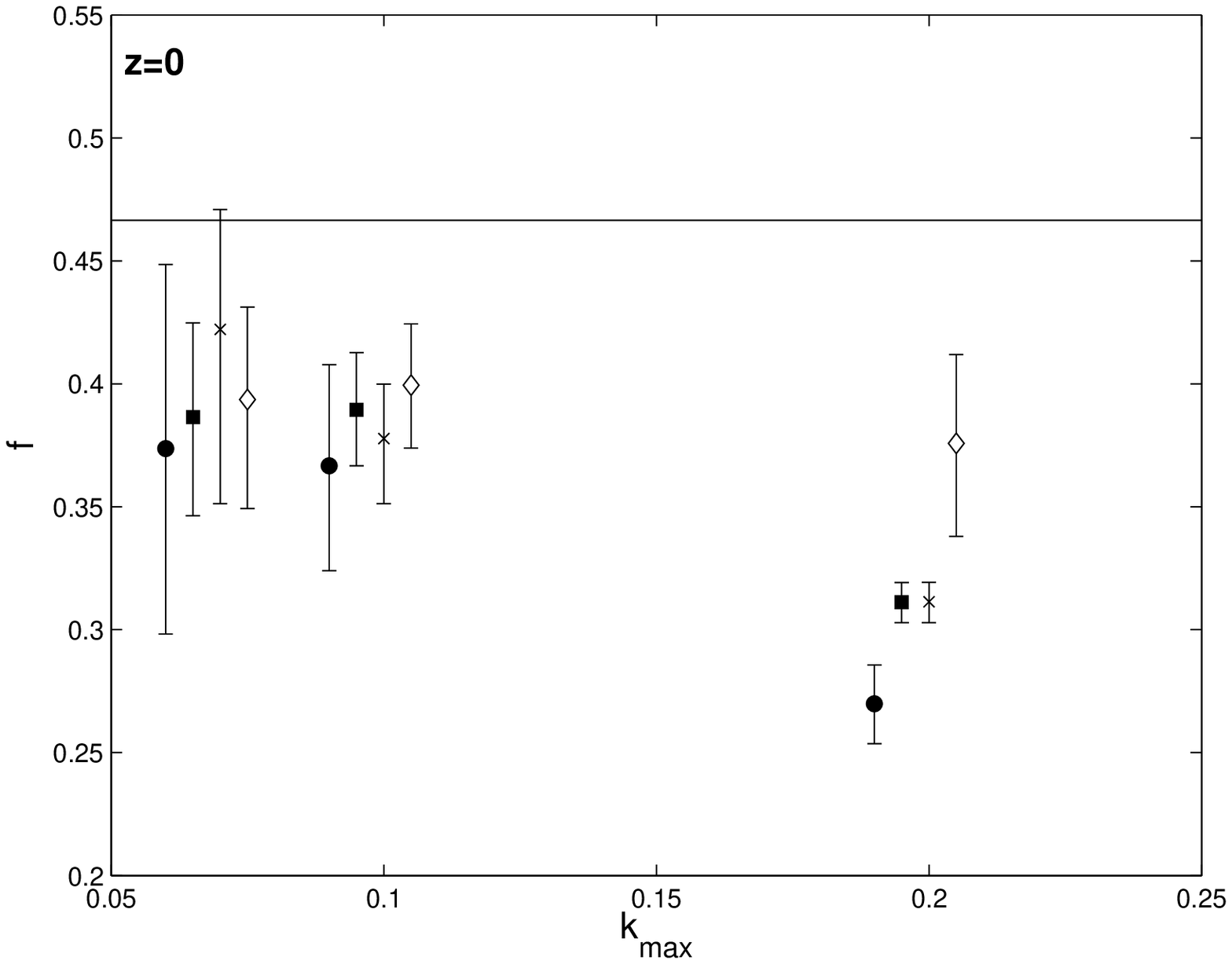}
\includegraphics[width = \columnwidth]{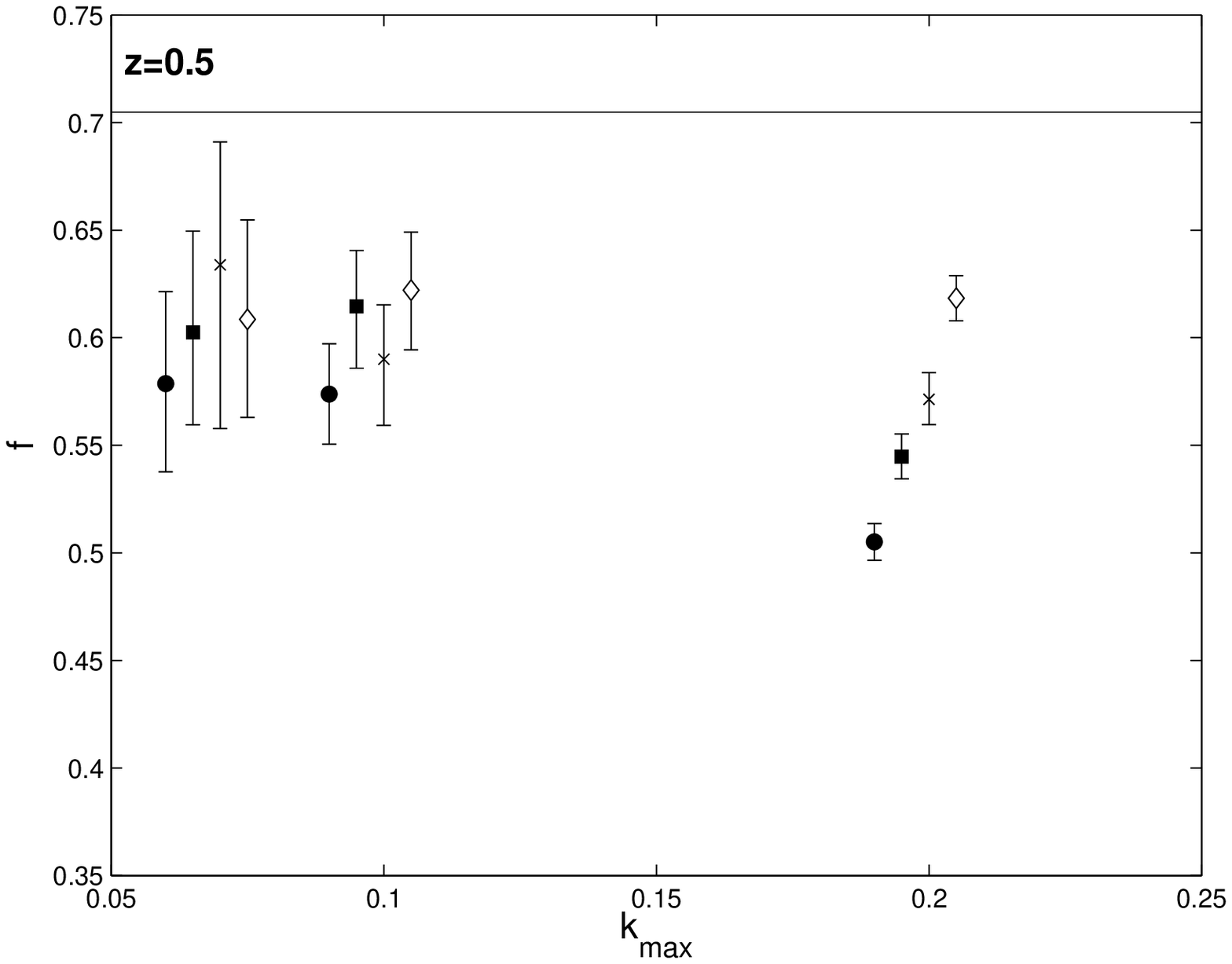}
\includegraphics[width = \columnwidth]{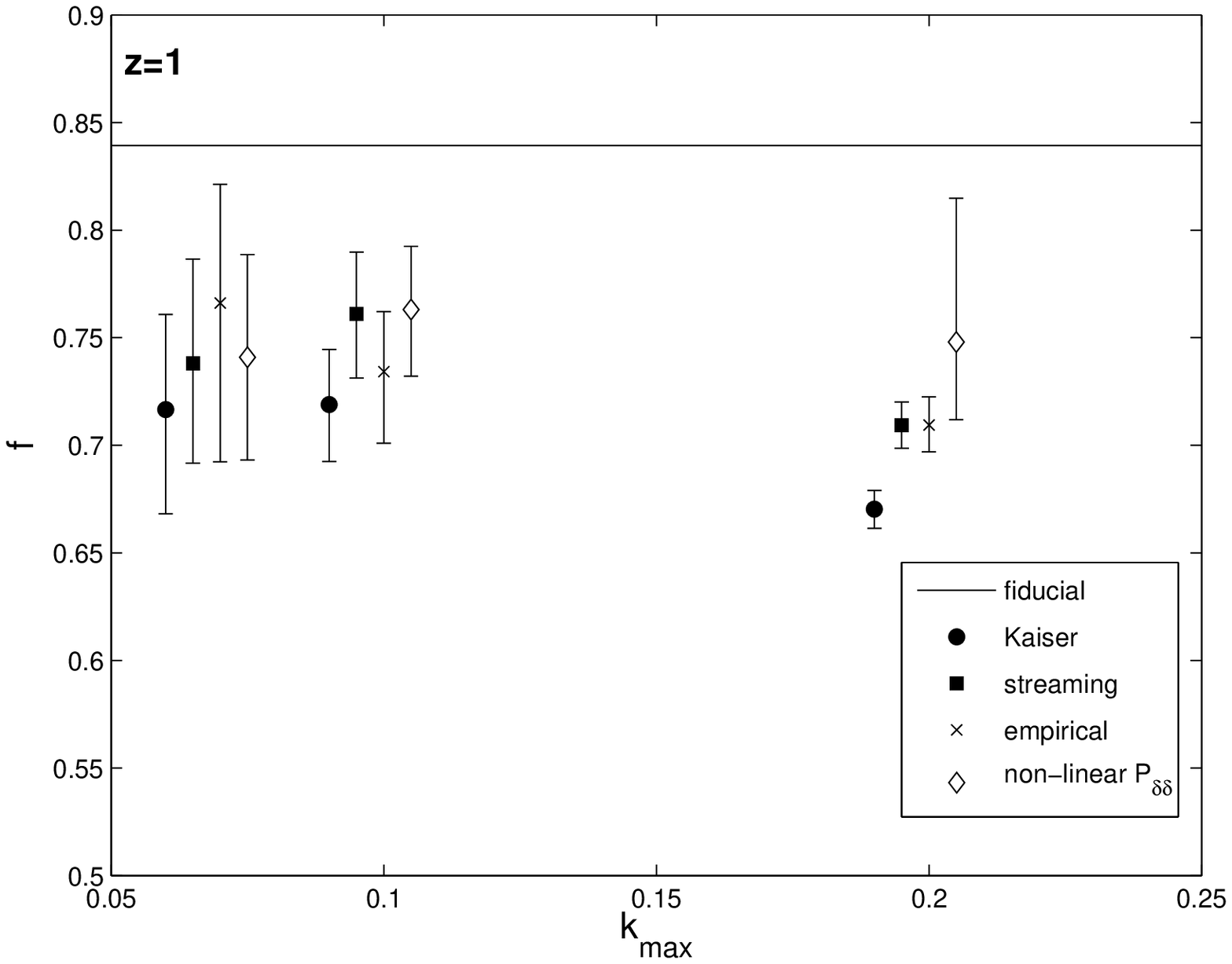}
\caption{Constraints on the growth rate using the various Kaiser-type
  models of the redshift space power spectrum evaluated at $z=0$
  (top), $z=0.5$ (middle), and $z=1$ (bottom), with cutoffs at
  k$_{max} = 0.07, 0.1, 0.2\,h$/Mpc. All the models are biased; the
  horizontal lines show the true value of $f$ at each redshift.  The
  fitted parameters are $\{f, b\}$, and $\sigma_v$ in the case of
  empirical damping. Note that some points have been offset by a small
  amount in k$_{max}$ for clarity.}
  \label{fig:kaiser_f_vs_kmax}
\end{center}
\end{figure}

Figure~\ref{fig:kaiser_f_vs_kmax} shows the marginalised $1\sigma$
confidence limits on the growth rate obtained from fits to the full
redshift space dark matter power spectrum in $\{k,\mu\}$ with 20 bins
in $\mu$ measured from an ensemble of N-body simulations at $z=0, 0.5,
1$. The only parameters that are allowed to vary are the growth rate
$f$ and the linear bias, $b$, while all others are kept fixed at their
fiducial value, with the exception of the empirical damping model
which has $\sigma_v$ as an additional free parameter. We have checked
the best fitting value of the growth rate remains the same within
1$\sigma$ when the linear bias is fixed to $b=1$ up to $k_{max}=0.1\,
h/$Mpc. However, at $k_{max}=0.2\,h/$Mpc, we found that the behaviour
was model dependent; sometimes this moved the fit closer to the
fiducial growth rate but in all circumstances, the 1$\sigma$ region
did not lie within the true value anyway.

We have quoted the mean likelihoods obtained from CosmoMC, rather than
the best fitting points, because these are considered to be more
robust against variations in chain length \citep{2002PhRvD..66j3511L},
although the differences are minor. The $1\sigma$ intervals are
calculated from the 1D minimum credible intervals obtained from the
posterior by using two tailed equal likelihood limits. We consider
four different models for the redshift space power spectrum: the
Kaiser limit (Eq.~\ref{Kaiser}), the streaming model
(Eq.~\ref{streaming}), the streaming model with non-linear
$P_{\delta\delta}$ and the empirical model which has the same form of
exponential damping as the streaming model, but $\sigma_v$ is treated
as a free parameter. These models and their free parameters are listed
in the top third of Table~\ref{modelnames}.

Each of the four models was fitted to a redshift space power spectrum
measured from N-body simulations with three cuts in $k$, namely
$k_{max}=0.07, 0.1, 0.2\,h$/Mpc. For each of these power spectra,
$k_{min} = 0.03\,h$/Mpc, since the larger modes have been contaminated
by the finite box size. This large scale cutoff was determined by
comparing the real space power spectrum measured from the simulations
to a reference power spectrum spectrum supplied by the CosmicEmu
package~\citep{2010ApJ...713.1322L}. We have also performed additional
convergence tests in redshift space at an extremely high redshift of
$z=19$, to confirm that on these scales we could recover the Kaiser
limit (Eqn~\ref{Kaiser}) with our N-body redshift space power
spectrum.

The performances of these Kaiser-type models are unimpressive
regardless of scale and redshift, even for $k_{max} = 0.07\,h$/Mpc at
$z=1$ where it is commonly thought that structure formation might be
sufficiently linear that the Kaiser model might be appropriate. While
the failure of the Kaiser limit is broadly consistent with the
findings of \cite{2011ApJ...726....5O, 2011ApJ...727L...9J}, in that
the growth rate is under predicted by these models, it is a little
surprising that neither the damping term nor the inclusion of the
non-linear power spectrum affects the best fitting value for the
growth rate. All these variants of the Kaiser limit assume that
$P_{\theta\theta} = f^2 P_{\delta\delta}$ and $P_{\delta\theta} = -bf
P_{\delta\delta}$, and the bias in these models regardless of damping
suggests that neglecting to include the non-linearity of large scale
motions by their simplistic relationship between density and velocity,
rather than the smearing of power from FoG effects, is the most
important systematic at low
redshifts~\citep{2004PhRvD..70h3007S,2011MNRAS.410.2081J}. But it is
not until we examine the Scoccimarro and Taruya$^{++}$ models in
Sections~\ref{sec:quasi} and~\ref{sec:morept} that we can see the
significance of the coupling between the density and velocity fields;
the extra terms in the Taruya$^{++}$ formalism affects scales as large
as $k_{max} = 0.07\,h$/Mpc.

\begin{figure}
\begin{center}
\includegraphics[width=\columnwidth]{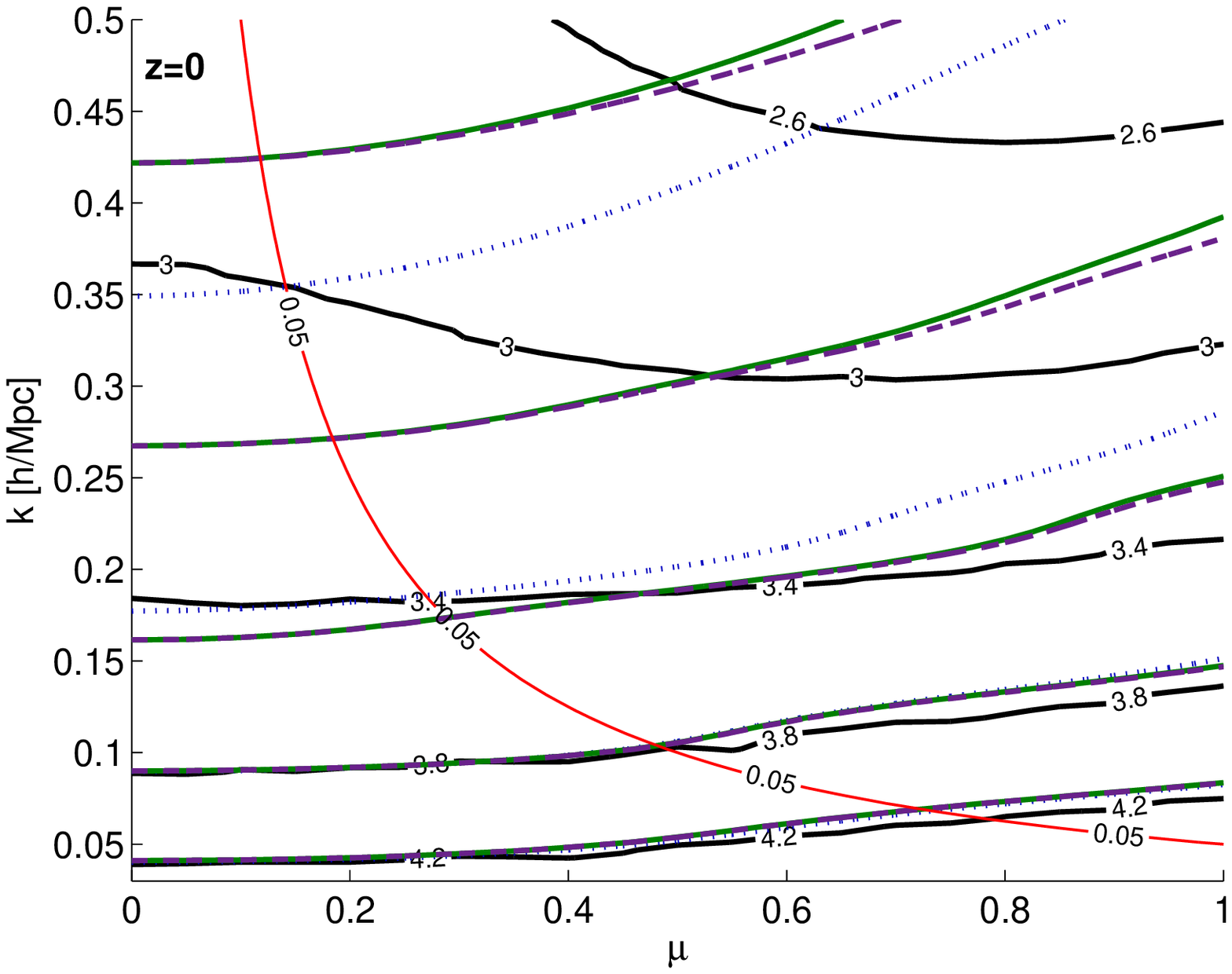}
\includegraphics[width=\columnwidth]{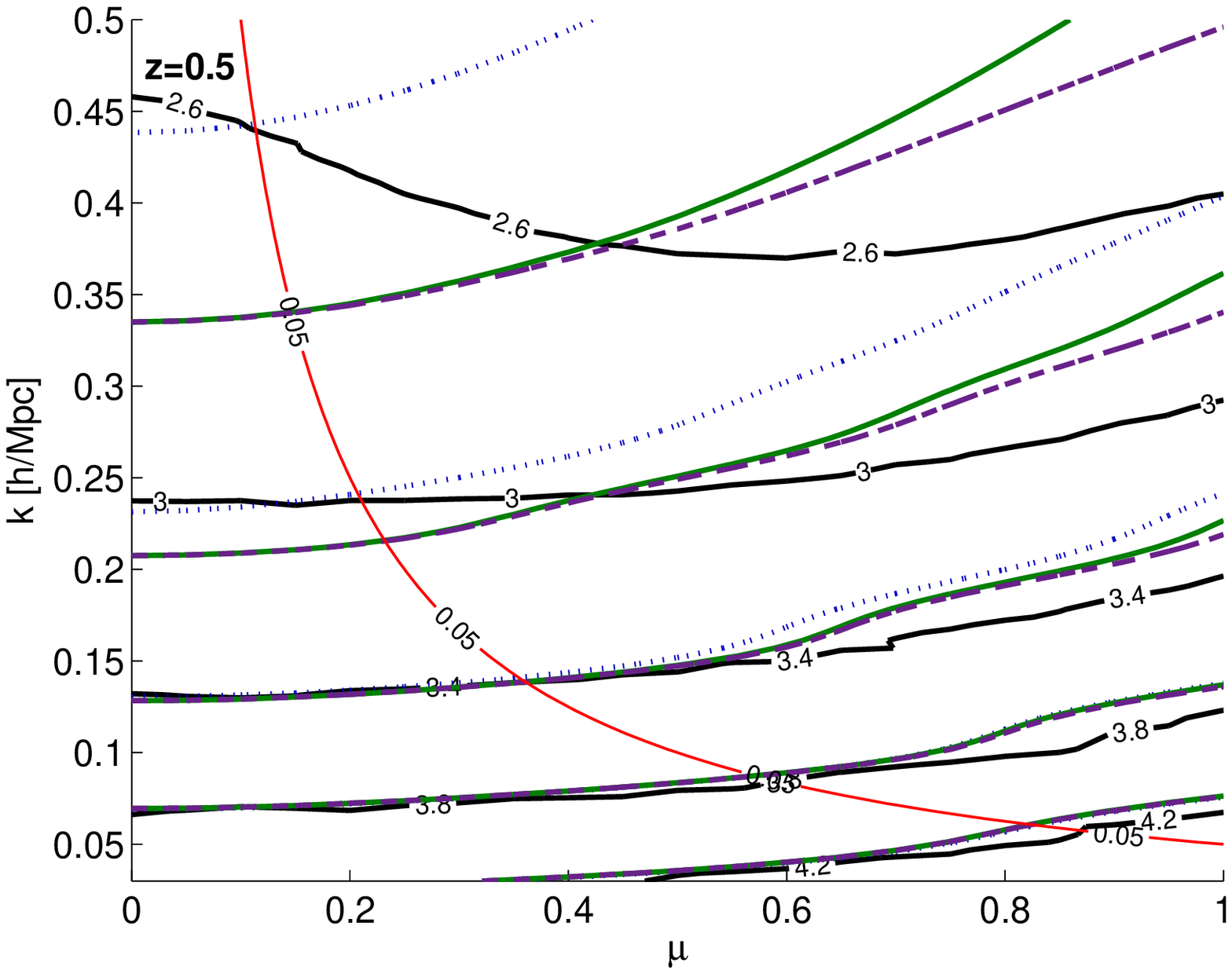}
\includegraphics[width=\columnwidth]{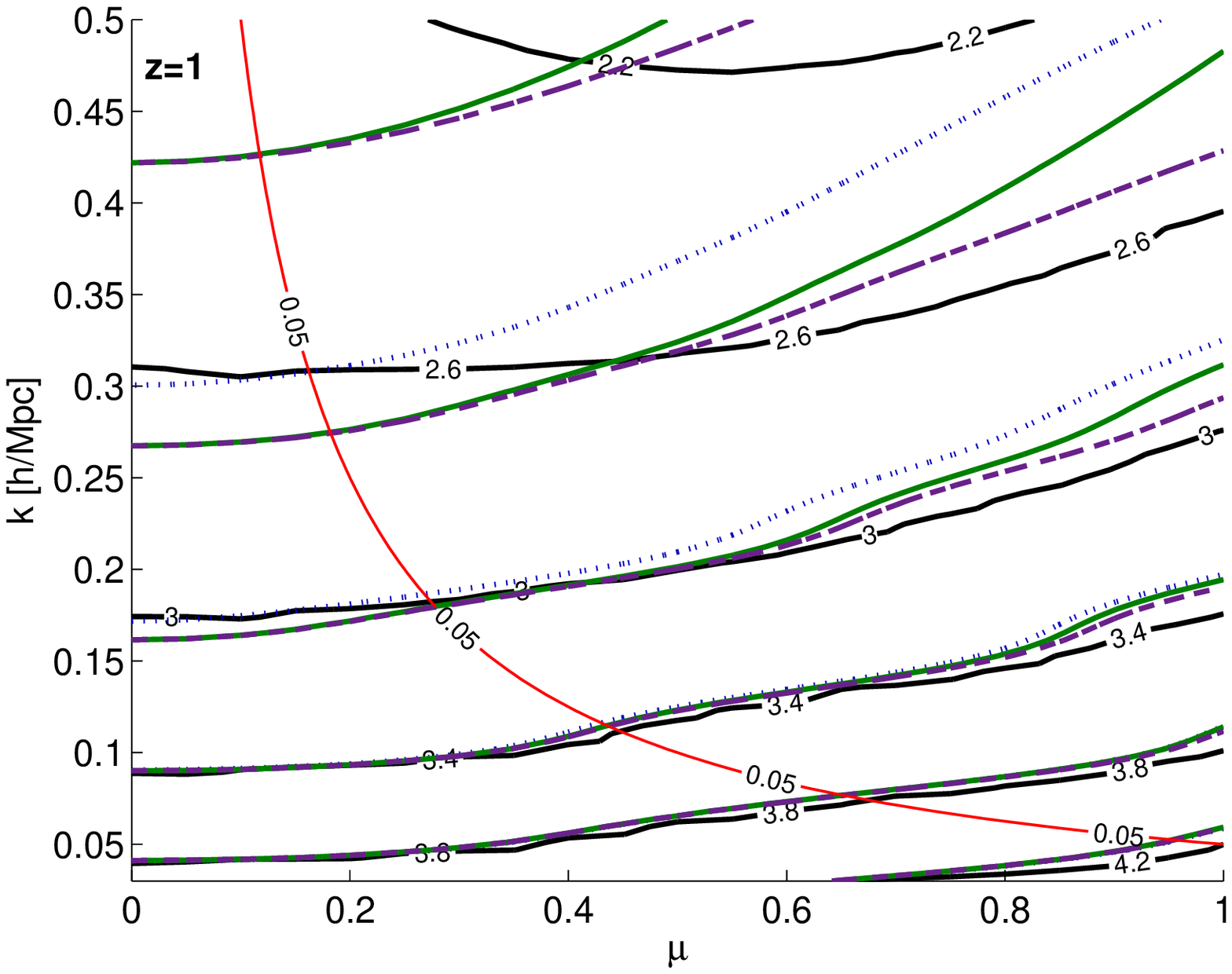}
\caption{$P^s(k,\mu)$ for $z=0,0.5,1$ calculated using the 
Kaiser limit without damping (dashed purple), the 
streaming model (solid green), and the streaming model using the 
  non-linear matter power spectrum (dotted blue), compared to measured
  redshift space power spectra from N-body simulations (black). The
  numbers on the black curves indicate the contour levels in
  $\log{P^s(k,\mu)}$. The hyperbolic red contours show constant
  $k\mu=0.05\,h$/Mpc.  FoG effects are manifested in the amplification
  of the black contours at large $k$ perpendicular to the line of
  sight ($\mu\approx0$). 
}
\label{fig:kaiser2D}
\end{center}
\end{figure}

Even though the streaming model accounts for the coupling between the
density and velocity fields with the factor of
$\exp\left(-f^2k^2\mu^2\sigma_v^2\right)$, the assumption that the
velocity distribution is Gaussian with a scale independent dispersion
term was shown to be of limited applicability by direct comparison to
N-body simulations in~\cite{2004PhRvD..70h3007S}. Neither does
exchanging the linear for the non-linear matter power spectrum in the
Kaiser formula work; this merely shifts the power spectrum in $k, \mu$
downwards to smaller $k$, while leaving the $\mu$ dependence the
same. In fact, it is the functional dependence on $\mu$ in these RSD
models that is inadequate, as suggested by the bias in the empirical
model, which is allowed to adjust for as much exponential damping as
required by the simulations. This angular dependence is a key point, 
and we explore it further in Sec.~\ref{sec:damp}.

Figure~\ref{fig:kaiser2D} shows the full 2D redshift space power
spectra of the models of this section (for simplicity we do not show
the empirical model) compared to the N-body power spectrum. We have
shown these as a function of $k$ and $\mu$, instead of in the
$k_{\perp}$--$k_{\parallel}$ plane, because we find it useful to fit
in terms of $k$ and $\mu$ so these figures provide a direct point of
comparison. None of these models are predicting the correct behaviour
at small angles (i.e.\ large $\mu$), even on large scales (i.e.\ small
$k$).  Looking at Fig.~\ref{fig:kaiser2D}, we can see that the
streaming and empirical damping models work in an average sense via a
happy coincidence: the linear theory matter power spectrum
underpredicts the amount of damping required at large $\mu$, but
happens to compensate for this by predicting less power at $\mu \sim
0$, i.e.\ the real space power spectrum is smaller than it ought to
be.

The inability of the Kaiser-type formulae to predict an adequate
amount of damping in the power spectrum as $\mu$ approaches unity
suggests that it may be worthwhile to consider truncating the power
spectrum in $k\mu$ instead of just scale alone. Unfortunately, we
found that for even quite conservative cuts such as $k\mu <
0.05\,h$/Mpc and $k < 0.2 \,h$/Mpc, no substantial improvements could
be gained: the growth rate is still biased by more than $1\sigma$
despite the increase in the error bars because of the weaker
dependence on $f$. Nonetheless, we would like to investigate the limit
at which $k$ and $k\mu$ can be truncated to reproduce the streaming
model with reasonable constraints on $f$, which doubles as a
consistency test. We found that the scales considered were
prohibitively large (no smaller than $k_{max} = 0.07 \,h$/Mpc could be
allowed) but the best fitting values of $f$ obtained were within
1$\sigma$ of the fiducial. In Fig.~\ref{fig:kaiser2D}, the hyperbolic
red curves show a contour of $k\mu$, and we find in
Sec.~\ref{sec:damp} that an accurate fitting function in terms of
$k\mu$ can allow us to extend robust extraction of the growth rate $f$
to higher $k$. Furthermore, in Sec.~\ref{sec:damp} we show that to
truly account for FoG non-linear effects, we must allow for a damping
term -- or rather an amplification term -- on small scales acting
perpendicularly to the line of sight.

\section{Quasi-linear models} \label{sec:quasi} 

As a next step, we investigate the Scoccimarro ansatz which presents a
simple extension to the Kaiser formula beyond linear theory using the
same techniques presented in Sec.~\ref{sec:kaiser}.  We use
Eq.~\ref{scoccimarro} to model the redshift space power spectrum and
extract the growth rate. To calculate $P_{\delta\delta}$,
$P_{\delta\theta}$ and $P_{\theta\theta}$, we use 1-loop SPT which was
shown to be accurate in $P_{\delta\delta}$ to 1\% upto $k=0.08\,h/$Mpc
at $z=0$ in real space by \cite{2009PhRvD..80d3531C} and a few percent
for $P_{\delta\theta}$ and $P_{\delta\delta}$ to $k=0.1\,h/$Mpc at the
same redshift~\citep[see Fig. 8 of][for a detailed comparison. Note
that the accuracy is comparable with improved PT]{2009PhRvD..80d3531C}.

\begin{figure}
\begin{center}
\includegraphics[width=\columnwidth]{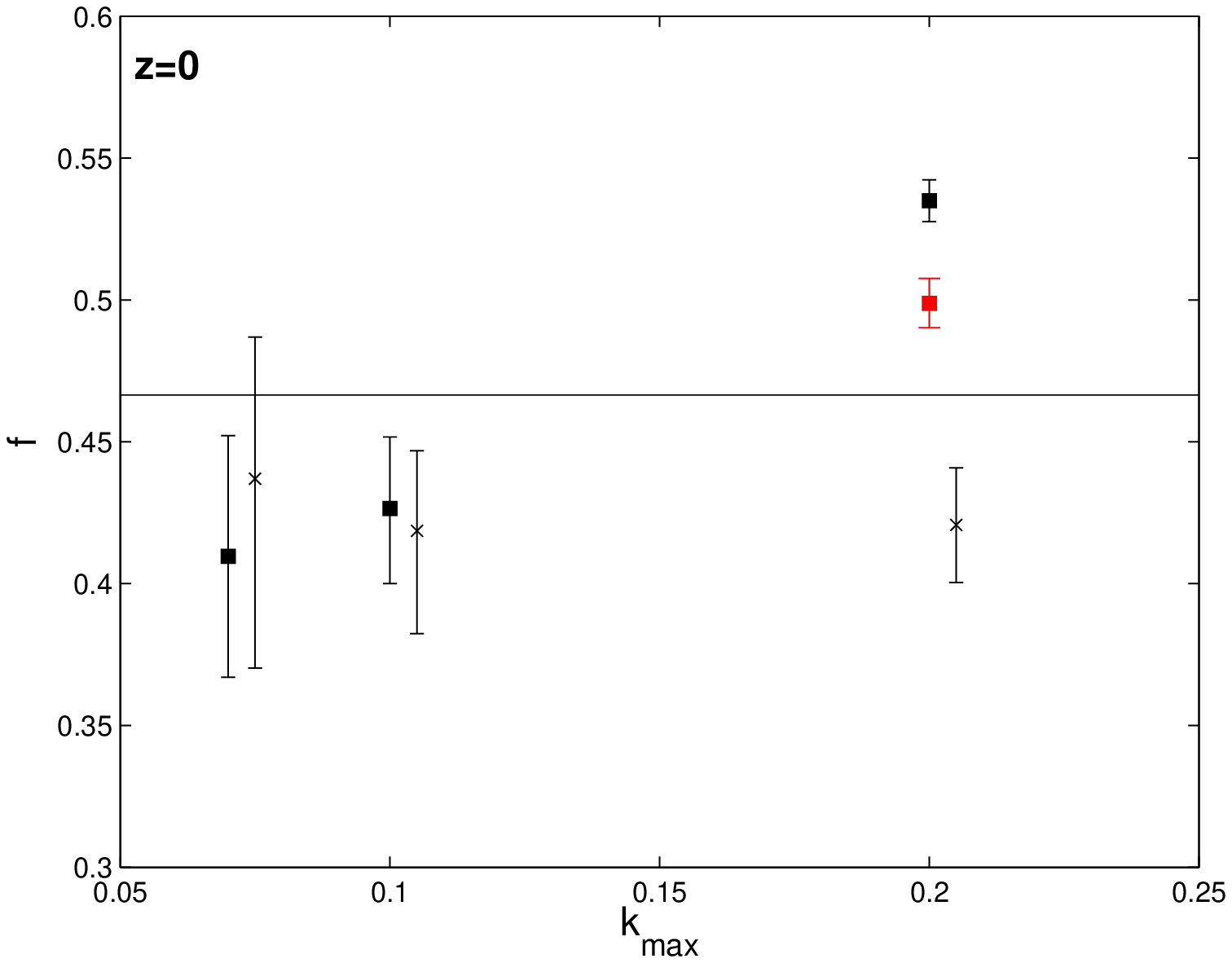}
\includegraphics[width=\columnwidth]{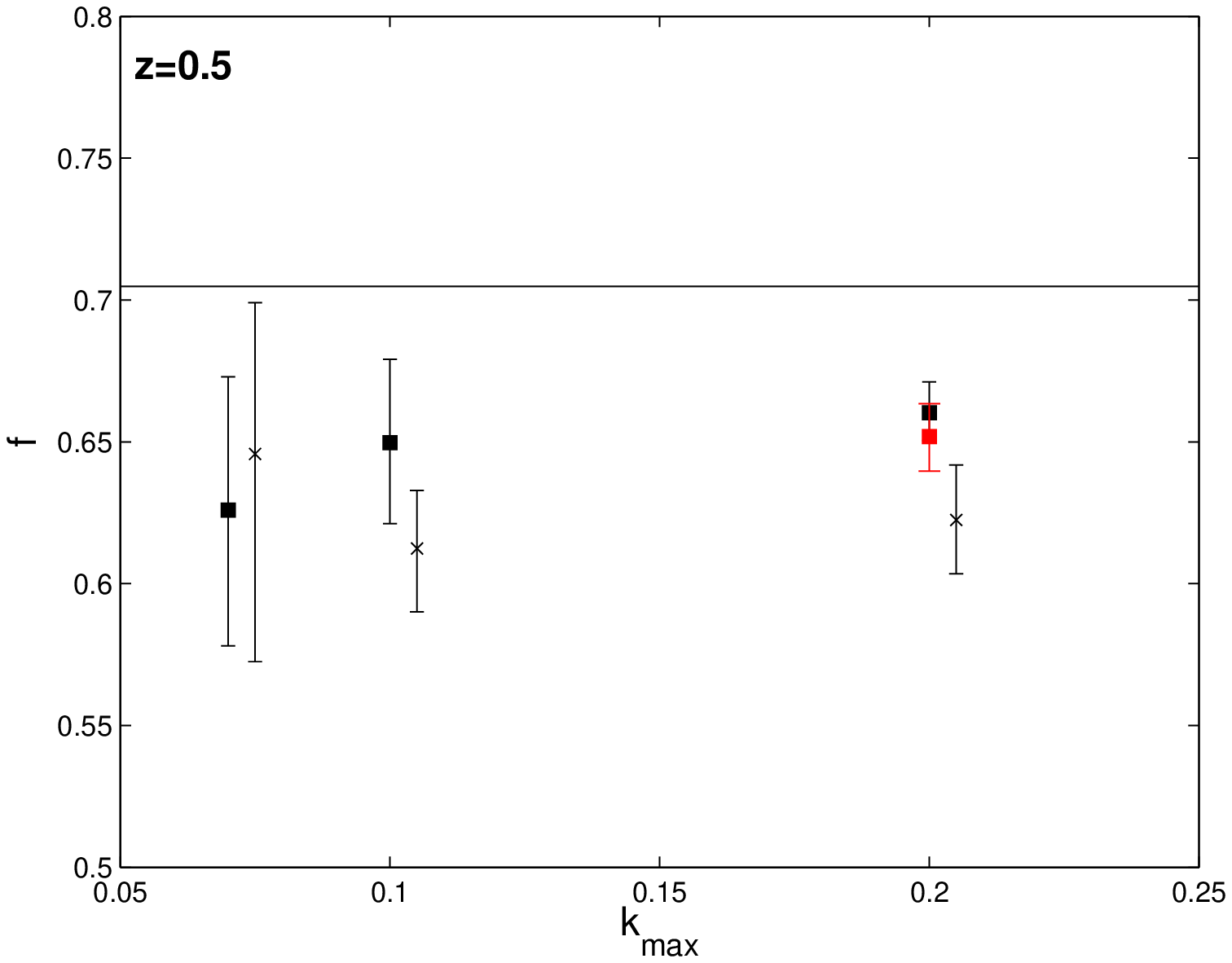}
\includegraphics[width=\columnwidth]{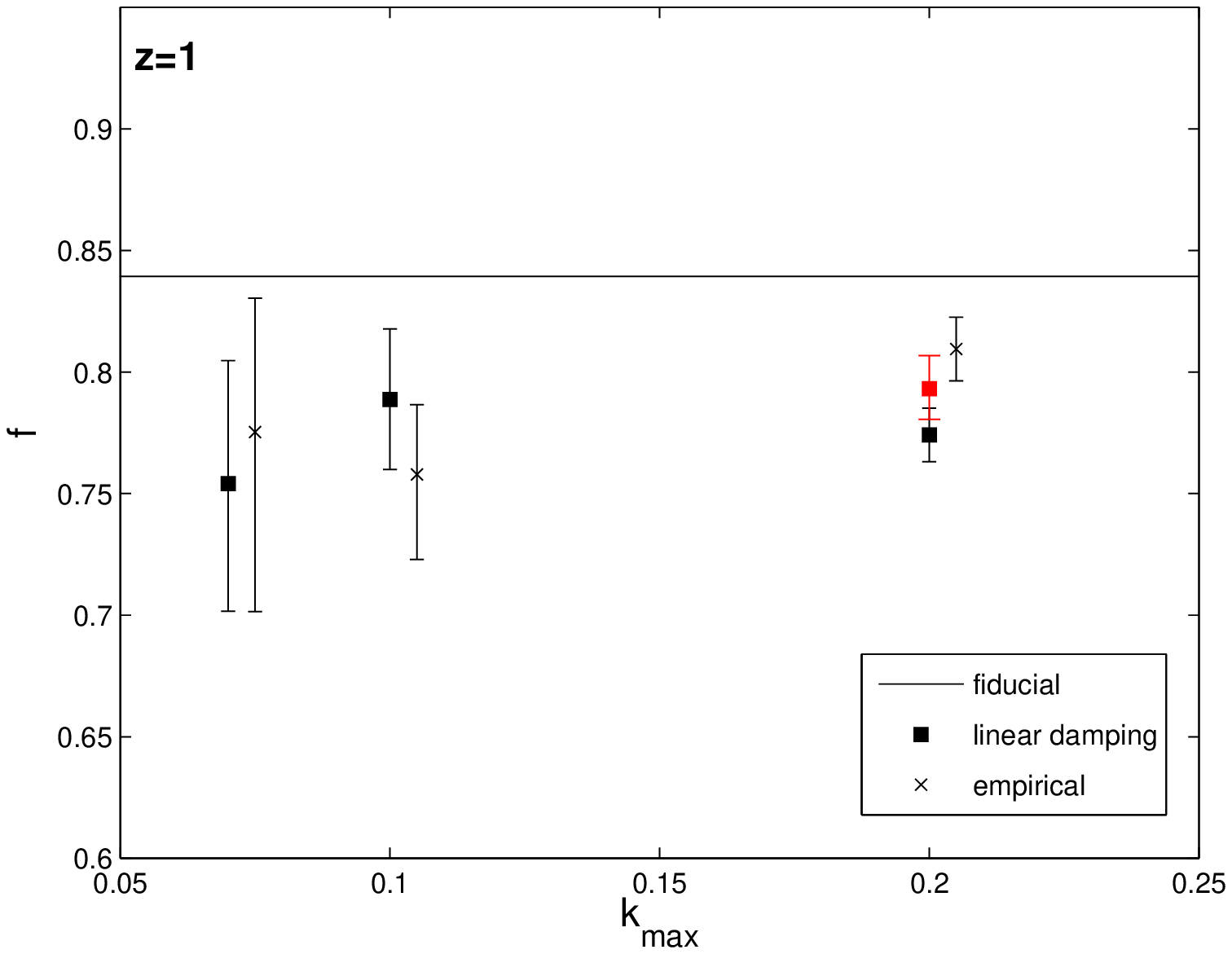}
\caption{As in Fig.~\ref{fig:kaiser_f_vs_kmax} but using the
  Scoccimarro ansatz with the real space non-linear power spectra
  evaluated using SPT. We consider two types of damping in these fits,
  exponential damping with $\sigma_v$ predicted by linear theory and
  allowing $\sigma_v$ to be a free, empirical parameter. The light red
  squares at $k_{max}=0.2 \,h$/Mpc are the result of fitting only over
  the range $0.1 \leq k< 0.2\,h$/Mpc. }
\label{fig:scoccimarro_f_vs_kmax}
\end{center}
\end{figure}

There is a minor improvement from the Kaiser formula to the
Scoccimarro results shown in Fig.~\ref{fig:scoccimarro_f_vs_kmax},
which is not sufficient to produce an unbiased result to
$1\sigma$. 
Even on large scales, such as for $k_{max}<0.07\,h$/Mpc, there is still
a small difference between the Kaiser limit and the Scoccimarro ansatz,
and both mean likelihoods are still skewed below the fiducial. This is
rather unlikely to be the effect of the size of the simulation because
the effect is model dependent, and changing the RSD model helps to
improve the fit as we shall see in Sec.~\ref{sec:morept}. 

Curiously, the Scoccimarro points in black seem to improve as the cut
in $k_{max}$ loosens, particularly at $z=0$ where the smaller scales
are weighted more strongly. The growth rate is being overestimated as
$k_{max}$ increases because the large scales underestimate $f$ while
the small scales tend to overestimate it. We can see this effect at
work in Fig.~\ref{fig:scoccimarro_f_vs_kmax}, in which we have also
explicitly split the $k$ range into two intervals, showing the high
range in red (the low range is given by the usual $k_{max}=0.1\,h$/Mpc
point). If the range of our fits to larger scales has an upper limit
in the range $0.1 < k_{max} < 0.2 \,h$/Mpc, then we could
coincidentally obtain a growth rate that is consistent with the
fiducial value. The interpretation is slightly more complicated with
the Scoccimarro model than in the Kaiser limit because there are two
competing effects; the real space power spectrum, $P_{\delta\delta}$,
is underestimated by SPT on small scales but the real space
$P_{\delta\theta}$ and $P_{\theta\theta}$ power spectra in SPT diverge
in comparison to the N-body result. The former requires a growth rate
to be larger than the fiducial value for the model to remain a good
fit, while the latter favours a smaller value for the growth rate.
The inconsistencies in the Scoccimarro ansatz are thrown into greater
relief when we consider what the red points are telling us: fitting
over the small and large scales jointly is not the same as fitting the
small and large scales independently. A more familiar setting for this
effect is the combination of two data sets, perhaps from different
cosmological probes, that are inconsistent -- their combined posterior
distribution need not overlap a similar region in parameter space as
their individual posteriors.

In addition, the use of empirical damping is unnecessary; in terms of
$1\sigma$ confidence levels, the two models are indistinguishable at
$k_{max} < 0.1 \,h$/Mpc and both models are equally poor on smaller
scales. Neither of these two models are able to consistently predict
the correct value of the growth rate and both give a value that is
substantially below the fiducial. This occurs because the model power
spectrum does not predict enough damping on large scales particularly
where the line of sight contribution is greatest; that is the
non-linear redshift space power spectrum has a lower amplitude than
expected as $\mu \rightarrow 1$, but on small scales it is the
reverse: FoG effects produce too much power perpendicular to the line
of sight. This is explored in greater detail in Sec.~\ref{sec:damp}.

\begin{figure}
\begin{center}
\includegraphics[width=\columnwidth]{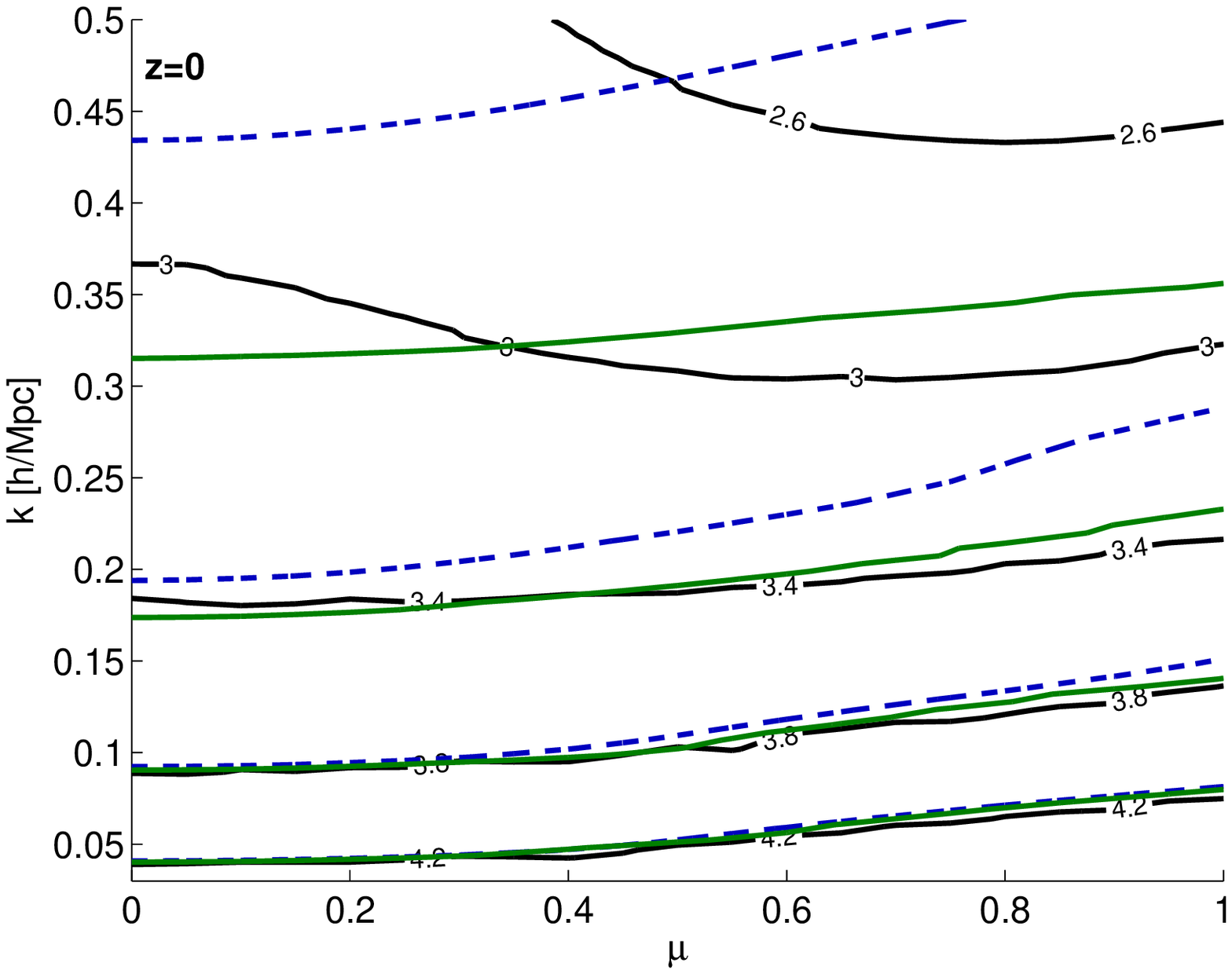}
\includegraphics[width=\columnwidth]{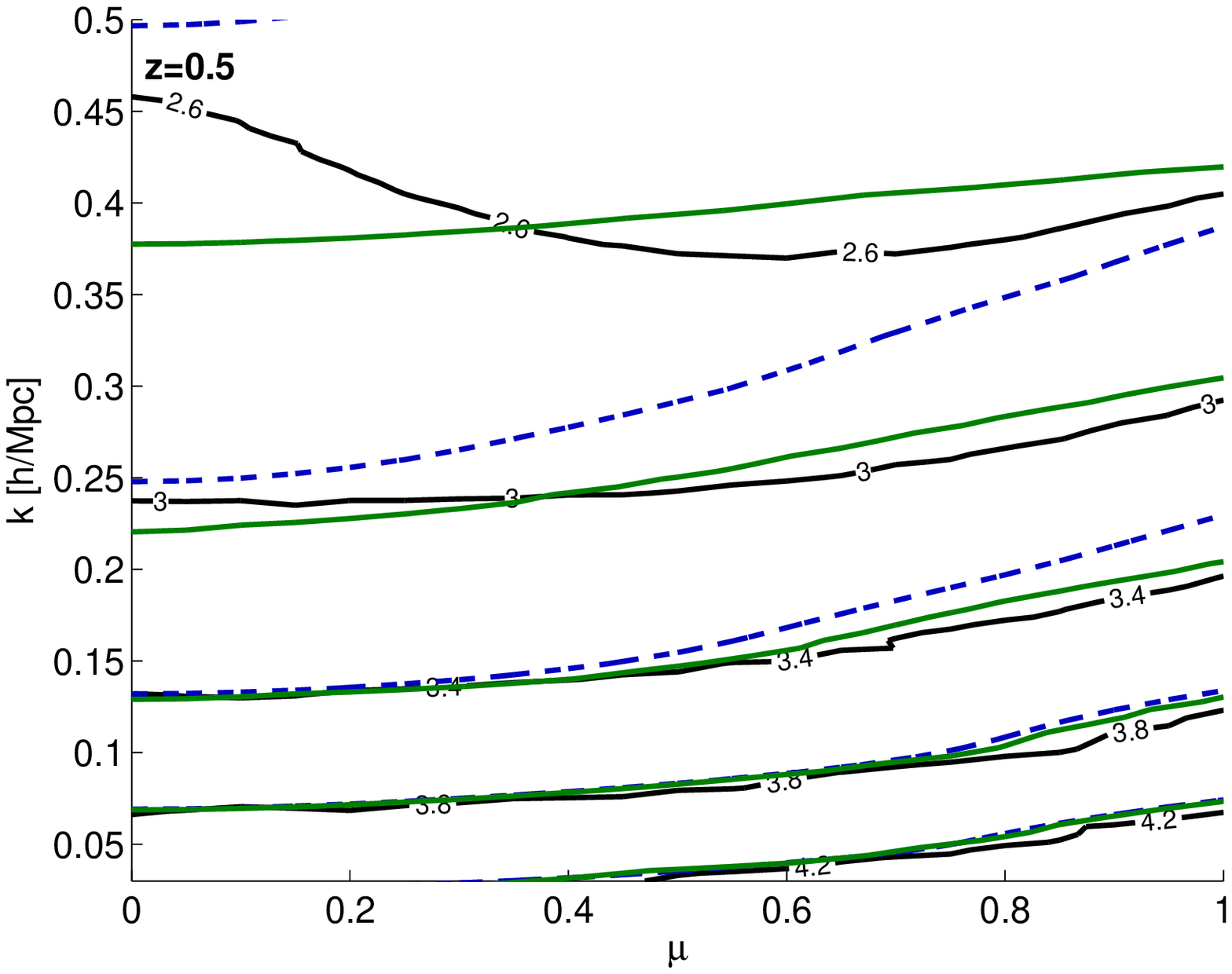}
\includegraphics[width=\columnwidth]{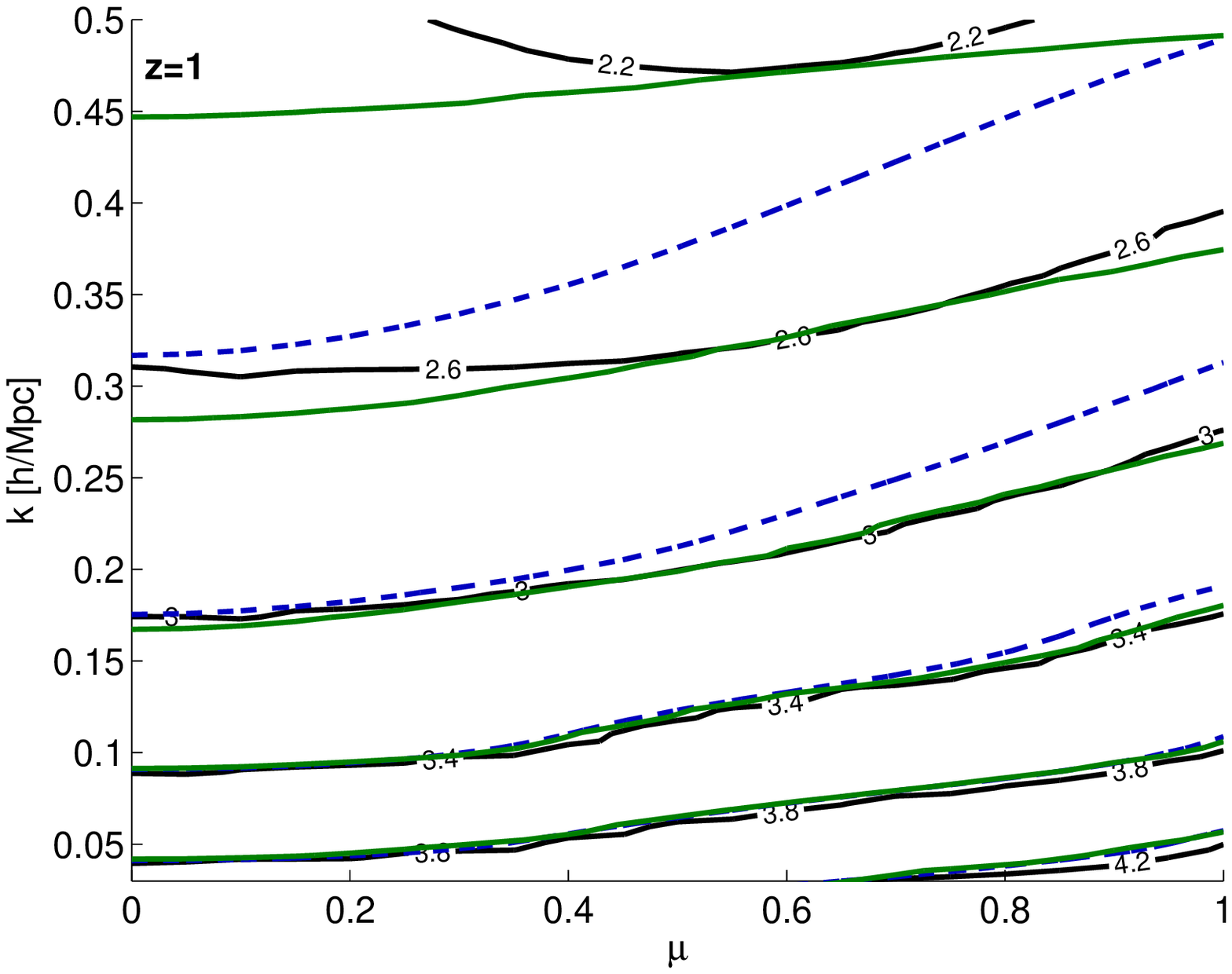}
\caption{$P^s(k,\mu)$ for $z=0,0.5,1$ as in Fig.~\ref{fig:kaiser2D}
but calculated using the Scoccimarro formula with linear theory
damping, and $P_{\theta\theta}$, $P_{\theta\delta}$ and
$P_{\delta\delta}$ measured from simulations (green) or 1-loop SPT
(dashed blue), compared to the redshift space power spectrum from
N-body simulations (black, labelled contours). }
\label{fig:scoccimarro2D}
\end{center}
\end{figure}

Compared to the simulations of \cite{2011ApJ...727L...9J}, we find two
areas of departure.  Although our simulations have the same box size
as theirs, the effective volume of our simulations obtained by
averaging the power spectra over the full ensemble of realisations is
substantially larger.  For the complete set of simulations, we found
weaker constraints on $f$ and more biased fits for the Scoccimarro
model.  However, by only using half of the simulations, the scatter
has increased such that we could obtain results that are consistent
with the fiducial cosmology for the Scoccimarro model. Perhaps a more
significant difference from \cite{2011ApJ...727L...9J} is that we have
used SPT for the real space power spectra instead of the fitting
function of~\cite{2011MNRAS.410.2081J}. Although this offers a
convenient alternative to perturbation theory, its accuracy is limited
to scales larger than $k \approx 0.1\,h$/Mpc because of the method
used to measure $P_{\theta\theta}$. We elaborate on this fitting
function in future work.

If we switch to using the exact forms for $P_{\delta\delta}$,
$P_{\delta\theta}$ and $P_{\theta\theta}$ as measured from the suite
of N-body simulations, we can separate the effects of the two
approximations that are involved, since we are no longer dependent on
the accuracy of SPT for the non-linear power spectra. This model is
represented in Fig.~\ref{fig:scoccimarro2D} by the green contours,
showing the redshift space power spectrum calculated with the
Scoccimarro formula using linear theory damping but
$P_{\delta\delta}$, $P_{\delta\theta}$ and $P_{\theta\theta}$ are
measured directly from the N-body simulations. This is what would be
produced by SPT if it were perfectly accurate up to $k \approx
0.2\,h$/Mpc, but we caution that there will be a deviation from the
true behaviour of $P^s(k,\mu)$ on smaller scales because the
resolution of the FFT grid is limited by the sparseness of the
particles when calculating $P_{\theta\theta}$.  Without the simulation
results being available for every model, the results of the
Scoccimarro ansatz could be improved over SPT by using a better
approximation to calculate the real space power spectra such as RPT or
a fitting formula such as~\cite{2011MNRAS.410.2081J}. One might be
tempted to attribute the inadequacies of the Scoccimarro formula to
the SPT power spectra, but this is surely not the case at large scales
where SPT is valid to 1\%. An inspection of
Fig~\ref{fig:scoccimarro2D} shows that on these scales, there
is a distinct difference between the Scoccimarro ansatz with linear
theory damping and the N-body power spectrum even at $z=1$. The
unsatisfactory results obtained in this approach with SPT leads us to
explore further perturbative schemes in the next section.

\section{Limits of Perturbation Theory} \label{sec:morept} 

\begin{figure}
\begin{center}
\includegraphics[width=\columnwidth]{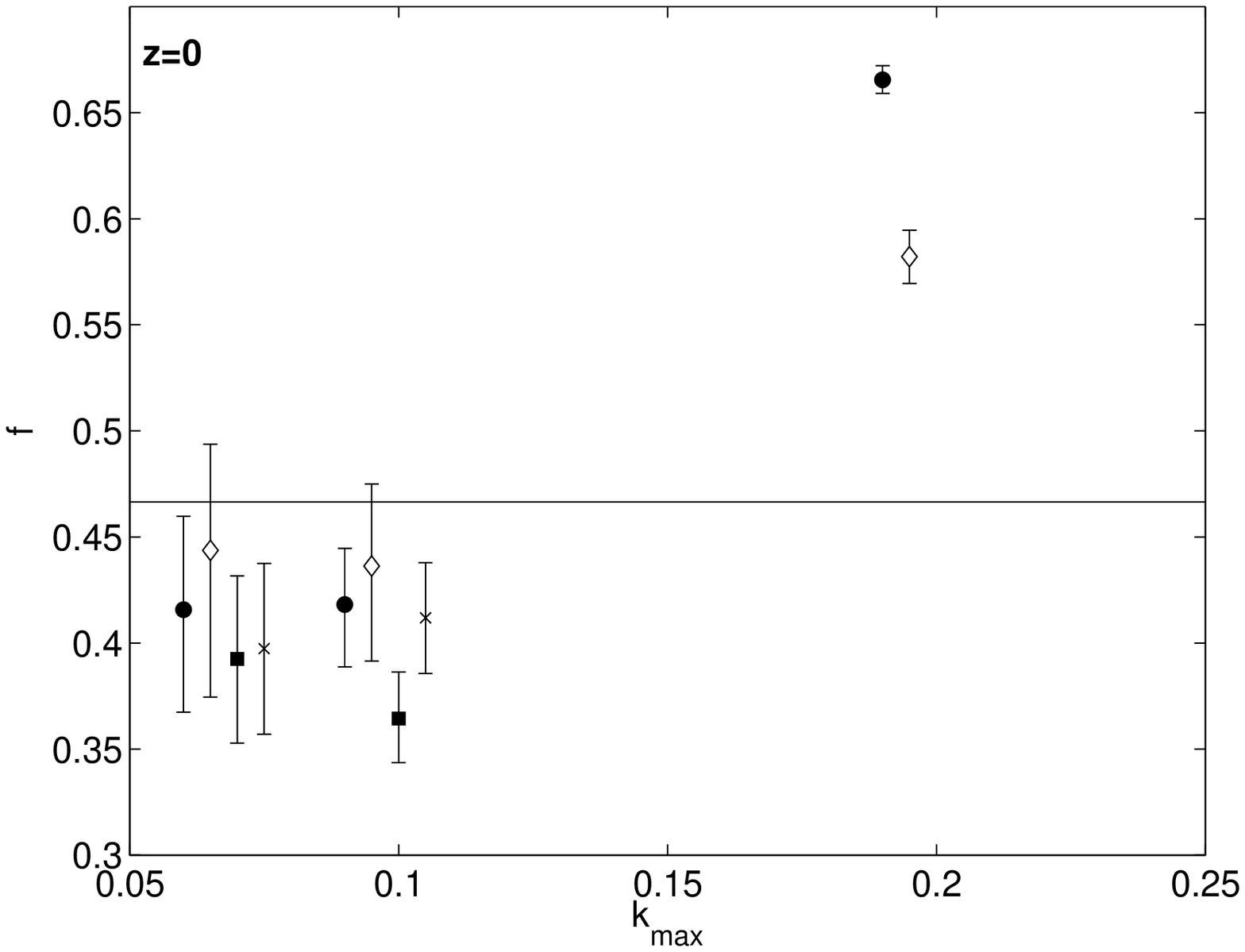}
\includegraphics[width=\columnwidth]{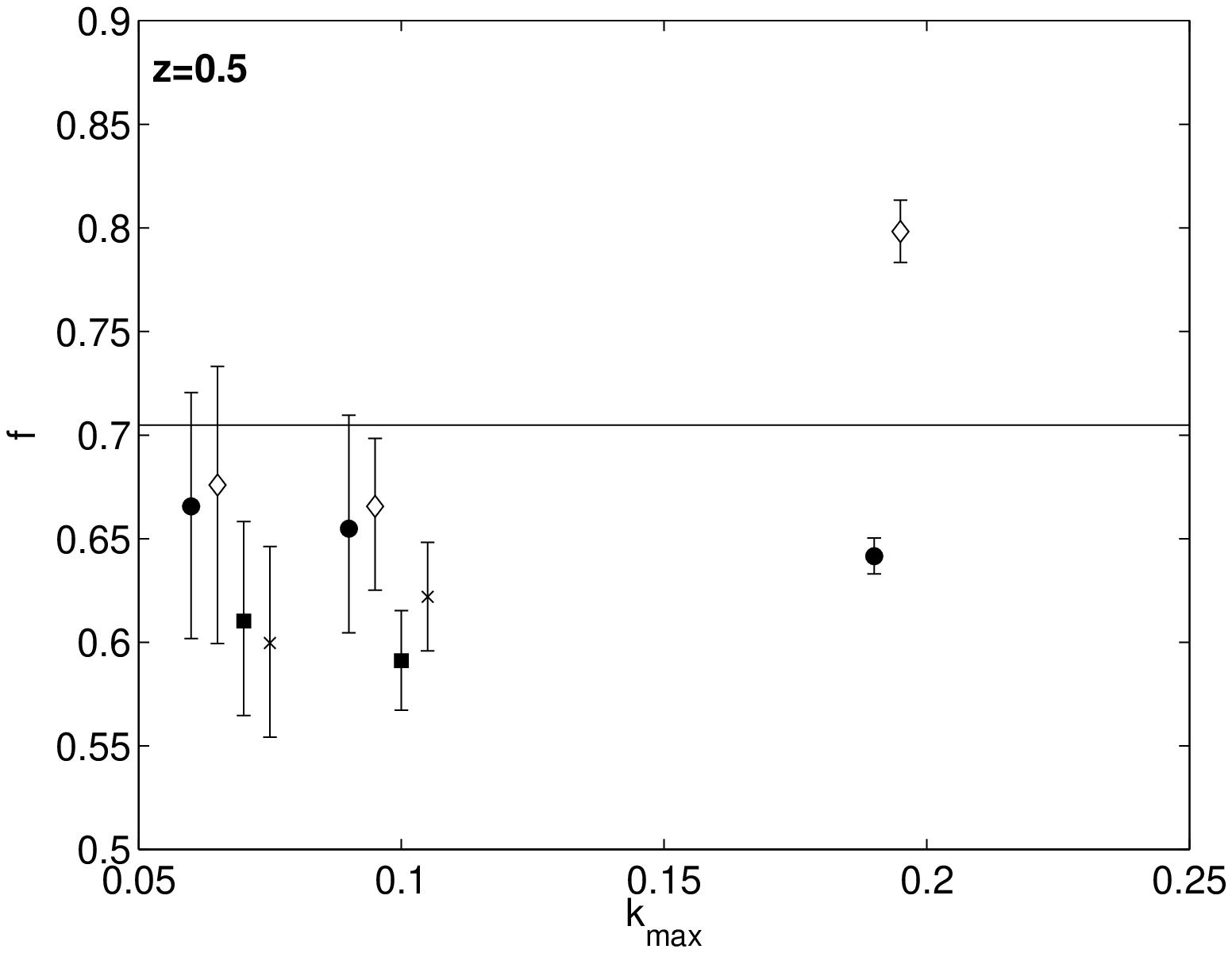}
\includegraphics[width=\columnwidth]{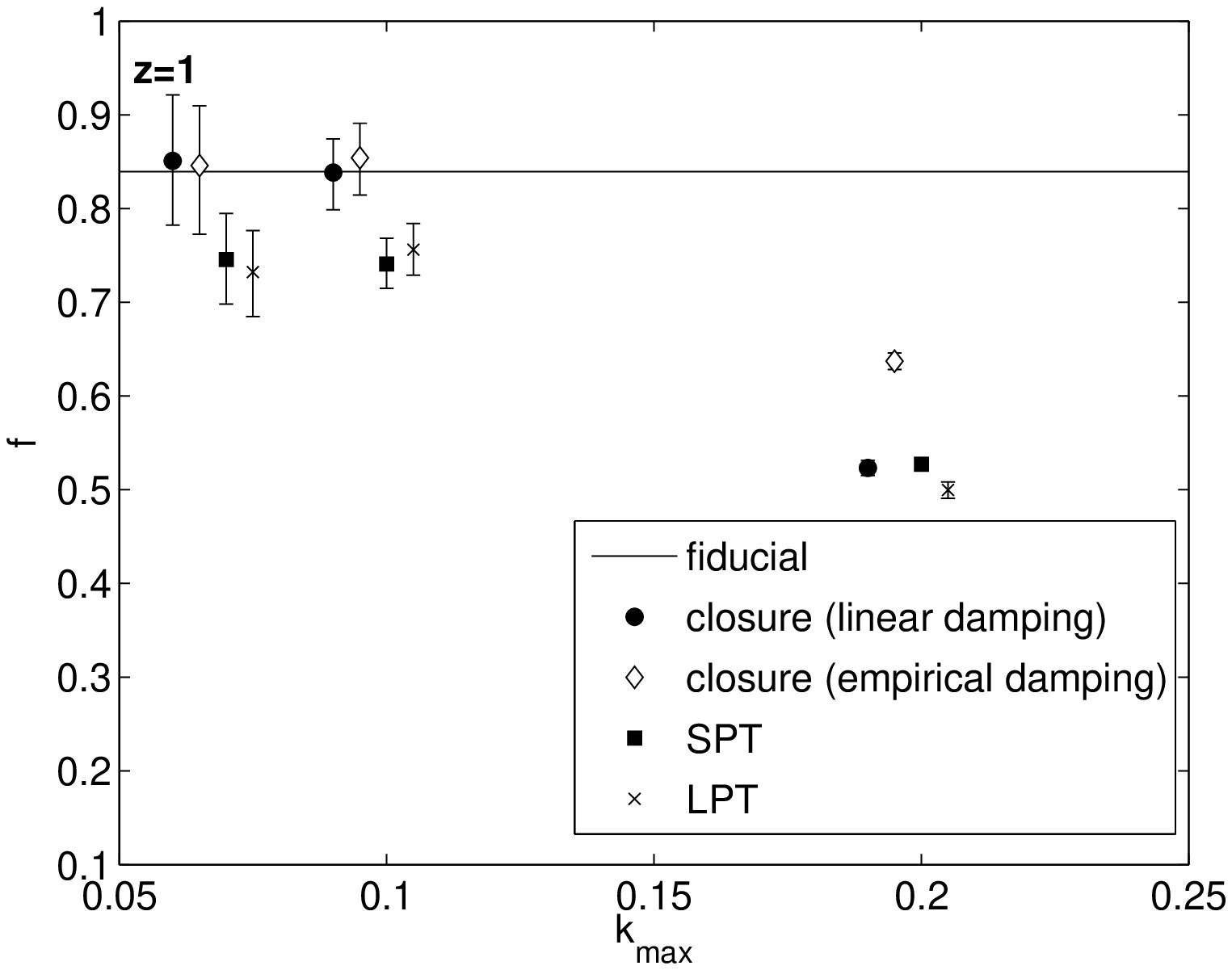}
\caption{1$\sigma$ constraints on the growth rates derived from
  modelling the redshift space power spectrum with SPT, LPT and
  Taruya$^{++}$ (with linear theory and empirical damping) models of
  the full redshift space power spectrum. We have considered the
  Taruya$^{++}$ model with both linear theory and empirical damping
  terms but the SPT model does not contain any damping. Values of $f$
  obtained for SPT and LPT have not been shown at $z=0,0.5$ for
  $k_{max}=0.2$ because they were consistent with zero.}
\label{fig:perturb_f_vs_kmax}
\end{center}
\end{figure}

We now consider more sophisticated models of redshift space
distortions, which are contained in the lower third of
Table~\ref{modelnames}. These are the SPT model proposed
by~\cite{1998MNRAS.301..797H,1999ApJ...517..531S} (Eq.~\ref{SPT}), the
LPT model of~\cite{2008PhRvD..77f3530M} (Eq.~\ref{LPT}) and model
of~\cite{2010PhRvD..82f3522T} (Eq.~\ref{closure}), which we refer to
as Taruya$^{++}$. These models rely on first performing an expansion
on the transformation from real to redshift space, and then the
density contrast is given by an additional perturbative scheme in real
space. To evaluate the terms in Eq.~\ref{closure}, we have used SPT to
obtain $P_{\delta\delta}, P_{\delta\theta}$ and $P_{\theta\theta}$ for
ease of comparison with the previous Scoccimarro fits. Unlike the
previous models that we have discussed, these models contain higher
order terms beyond $\mu^4$, up to $\mu^8$. The values of $f$ obtained
from using these perturbation theory approaches are shown in
Fig.~\ref{fig:perturb_f_vs_kmax}. Note that we calculate both the
corrections suggested by the Taruya$^{++}$ model and the
$P_{\delta\delta}, P_{\delta\theta}$ and $P_{\theta\theta}$ used in
Fig~\ref{fig:perturb_f_vs_kmax} using SPT.

\begin{figure}
\begin{center}
\includegraphics[width=\columnwidth]{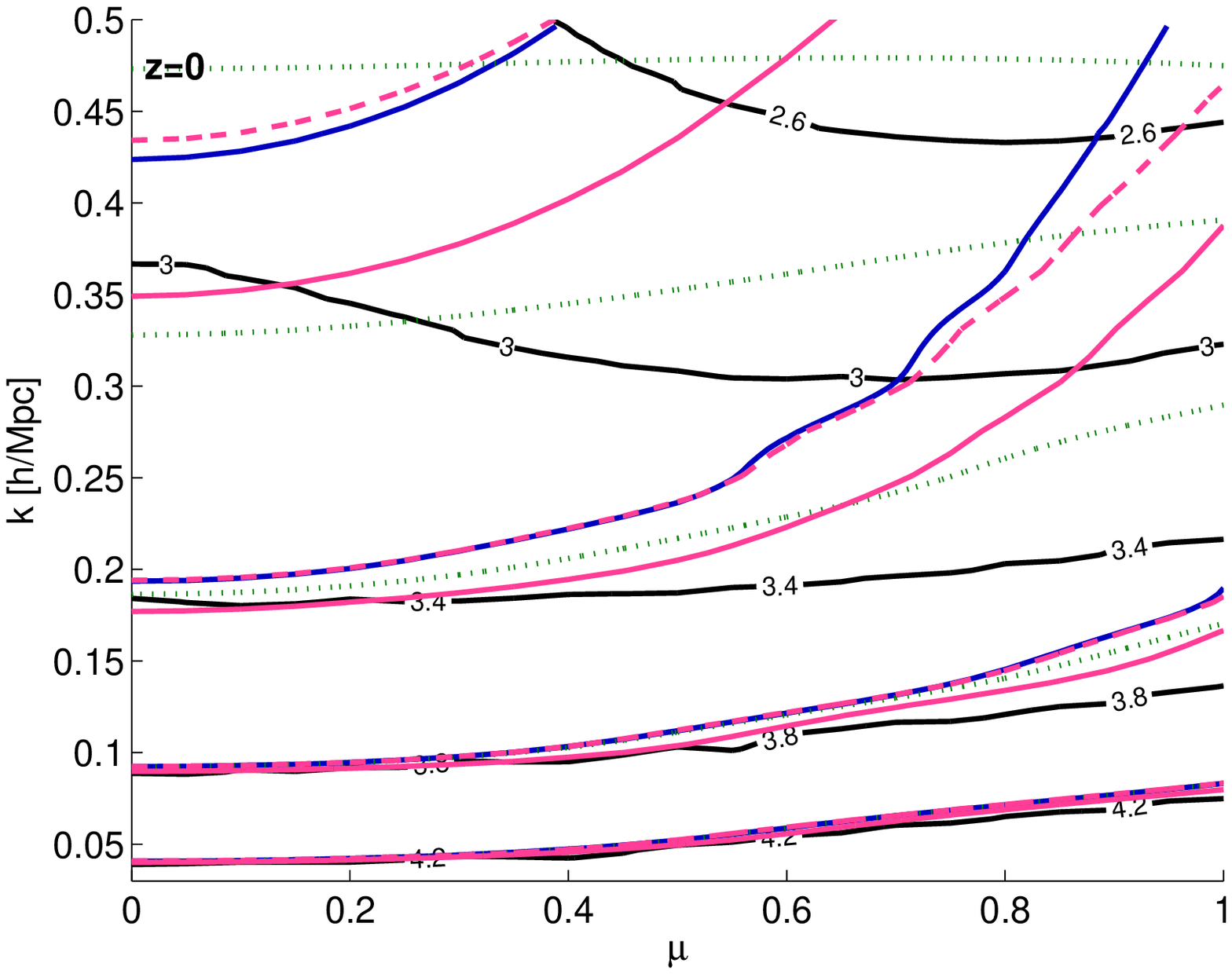}
\includegraphics[width=\columnwidth]{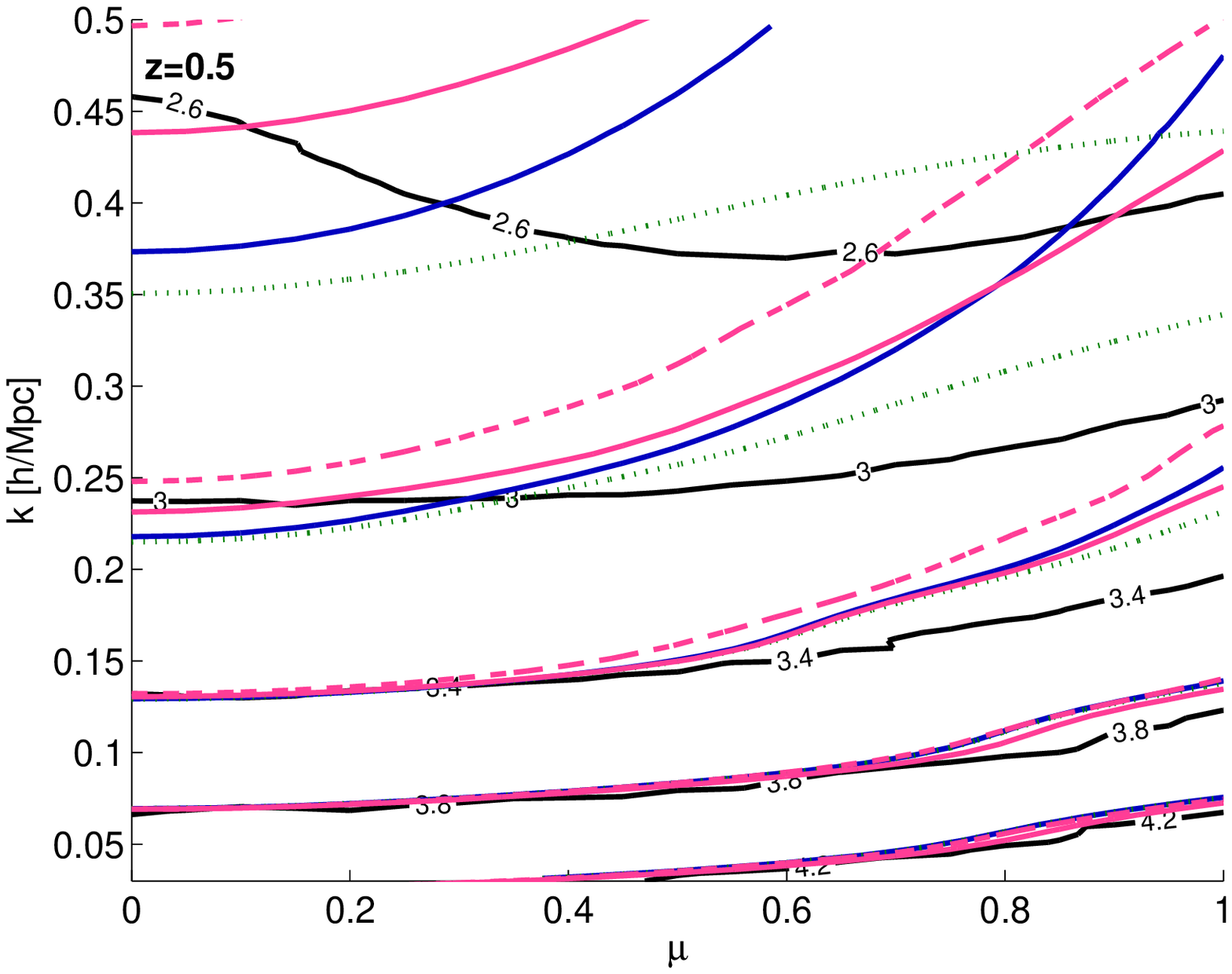}
\includegraphics[width=\columnwidth]{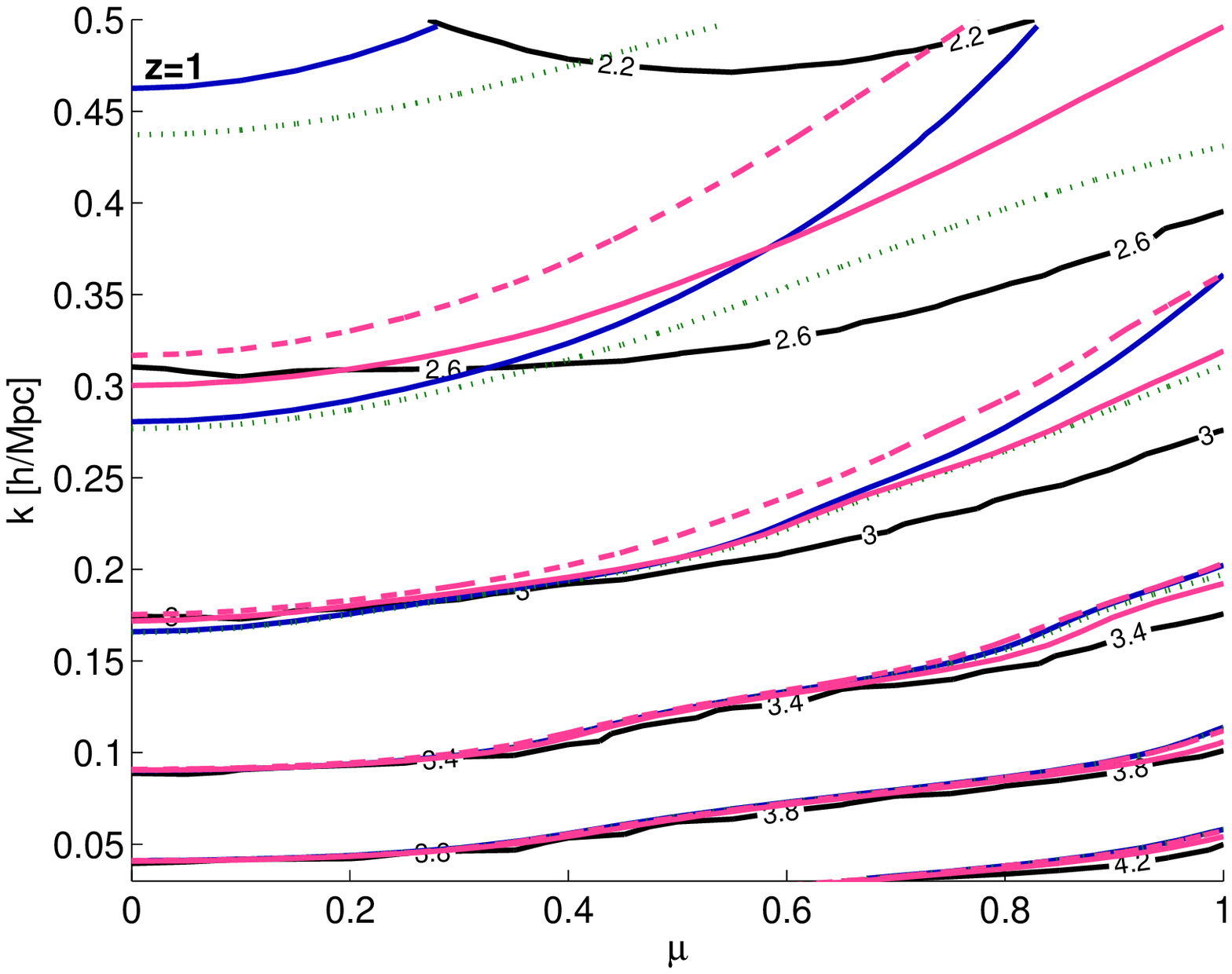}
\caption{$P^s(k,\mu)$ for $z=0,0.5,1$ as in Fig.~\ref{fig:kaiser2D},
  but using SPT (blue), LPT (dot-dashed green) and Taruya$^{++}$ (solid
  and dashed pink) treatments of the full redshift space power
  spectrum, and compared to N-body results (black, labelled). We plot
  the Taruya$^{++}$ power spectrum with two different methods of
  evaluating the terms: dashed pink curves correspond to using SPT for
  $P_{\delta\delta}, P_{\delta\theta}$ and $P_{\theta\theta}$ and
  solid pink curves use the exact N-body power spectra.  The SPT
  curves here do not include any damping terms and the Taruya$^{++}$
  curves use a linear theory damping term. }
\label{fig:perturb2D}
\end{center}
\end{figure}

Comparing Fig.~\ref{fig:scoccimarro_f_vs_kmax} and
\ref{fig:perturb_f_vs_kmax} (and the shapes of the power spectra in
Fig.~\ref{fig:scoccimarro2D} and~\ref{fig:perturb2D}), it is clear
that both LPT and SPT models of RSD perform substantially {\it
worse\/} than both the Kaiser and the Scoccimarro models. In fact, at
$z = 0$ and $z =0.5$ for $k_{max} > 0.1 \,h$/Mpc, we were unable to
obtain any constraints on $f$ using these models, as the $1\sigma$
interval was consistent with 0. Within $1\sigma$, the values of $f$
given by LPT and SPT are indistinguishable, but the shape of the LPT
power spectrum in Fig.~\ref{fig:perturb2D} is somewhat more
reasonable, despite the damping term in LPT having a stronger
dependency with $\mu$. The bias of the growth rate is partially due to
the strength of the constraints; the extra information contained in
the higher order terms means that the errors on the growth rate are
smaller. One of the advantages of considering terms beyond linear
theory is the additional information contained in those extra terms
that can break the degeneracy between $f$, $b$, and $\sigma_8$. For
instance, although the fits are more biased in SPT and LPT, $P^s_{22}$
and $P^s_{13}$ contain different combinations of $f$, $b$, and
$\sigma_8$ and some of these terms are proportional to
$(f\sigma_8)^2\sigma_8^2$ and $(b\sigma_8)^2\sigma_8^2$.

But by far the dominant source of the discrepancy is that there are
two levels of approximations that enter into the formalism: the
density perturbation expansion and the mapping from real to redshift
space.  For the SPT redshift space power spectrum, the transformation
from real to redshift space is expanded to third order and then the
real space overdensities are perturbed to $\delta^3$ for the 1-loop
terms~\citep{1998MNRAS.301..797H}. The derivation of the LPT redshift
space power spectrum proceeds in a rather different manner, but the
final expression can be related to SPT by expanding the linear theory
damping term that is predicted self consistently in the
theory~\citep{2008PhRvD..77f3530M}. The overall effect is to reduce
the accuracy of perturbation theory in redshift space relative to its
efficacy in real space. There are also several coincidental
occurrences that are advantageous to the Kaiser and Scoccimarro
formulae that are absent in the LPT and SPT redshift space power
spectra.

The SPT and LPT models in fact miss important cross terms between the
velocity and density fields that are included in the Taruya$^{++}$
expression, see Eq.~26 of~\cite{2010PhRvD..82f3522T}. For this reason,
the Taruya$^{++}$ power spectrum is able to model the redshift space
power spectrum with a smaller bias in the growth rate, as seen in
Fig.~\ref{fig:perturb_f_vs_kmax} on all scales and at all redshifts,
but we found that the posteriors on $b$ were biased beneath $b=1$.  It
is worth noting, however, that all the redshift space power spectra
considered in this section can be related to one another through the
addition of extra terms in the perturbation theory expansion. Much of
the variation in the estimates of the redshift space power spectrum
obtained from perturbation theory originates from a judicious choice
of terms to sum in the expansion of the density contrast.

We emphasise that for the dashed pink curves in
Fig.~\ref{fig:perturb2D} (and the results presented in
Fig.~\ref{fig:perturb_f_vs_kmax}) the SPT expansions of
$P_{\delta\delta}, P_{\delta\theta}$ and $P_{\theta\theta}$ were used
instead of the improved PT expressions and this contributes to the
deviation between our results and~\cite{2010PhRvD..82f3522T} who found
that their expression provides an unbiased description of the redshift
space power spectrum up to $k < 0.205\,h$/Mpc at $z=1$. Yet, a
comparison of Fig.~\ref{fig:scoccimarro2D} and
Fig.~\ref{fig:perturb2D} reveals that it is the Scoccimarro ansatz
that follows the N-body contours more faithfully than the
Taruya$^{++}$ model when using the exact non-linear expressions for
$P_{\delta\delta}, P_{\delta\theta}$ and $P_{\theta\theta}$ for both
models. The solid pink curves in Fig.~\ref{fig:perturb2D} that
correspond to the Taruya$^{++}$ model with the real space
$P_{\delta\delta}, P_{\delta\theta}$ and $P_{\theta\theta}$ measured
directly from N-body simulations are especially good near $\mu=0$ as
the Taruya$^{++}$ expression uses an expansion in $\mu$. None of the
models though get the angular dependence right at lower redshifts.

\section{Angular Dependence and Damping}\label{sec:damp}

\begin{figure}
\begin{center}
\includegraphics[width=\columnwidth]{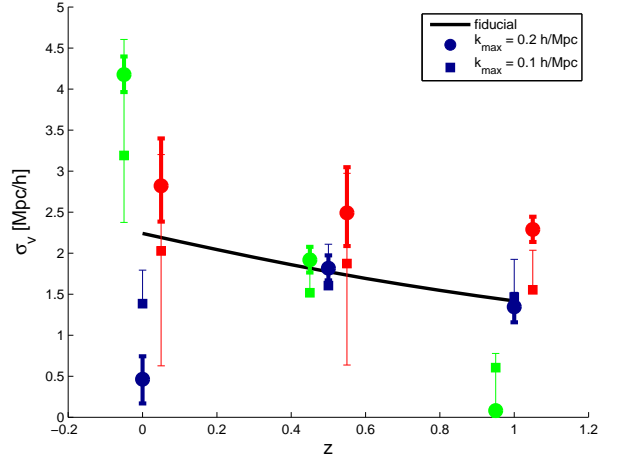}
\caption{Values of $\sigma_v$ determined from using empirical damping
  with the Kaiser limit (blue, centered) and the Scoccimarro formula
  (red, offset right) and the Taruya$^{++}$ model (light green, offset
  left) at $z = 0, 0.5, 1$ for $k_{max} = 0.1$ and $0.2\,h$/Mpc. For
  comparison, we have also shown the amount of damping predicted by
  linear theory (black curve). We have removed the lower error bars
  for those points whose 1$\sigma$ limit is consistent with zero.}
\label{fig:sigma_v}
\end{center}
\end{figure}

The damping term in the power spectrum arises from the velocities and
also non-linearities.  We first examine the values of $\sigma_v$
obtained when we allow it to vary as an extra parameter when fitting
for the growth rate. We allow for redshift variation between different
snapshots but have assumed that $\sigma_v$ is independent of
scale. Figure~\ref{fig:sigma_v} shows the $1\sigma$ constraints on
$\sigma_v$ for the Kaiser, Scoccimarro and Taruya$^{++}$ models with
and without the linear bias (recall that the amount of damping
required in LPT is predicted by the model) in comparison to the linear
theory prediction shown by the black curve. The bottom error bars have
been suppressed where they are consistent with $\sigma_v=0$.

For many of the models at most of the redshifts, $\sigma_v$ is not
helping to fit the power spectrum shape and so the values are
consistent with $\sigma_v=0$. Only the Scoccimarro model with
$k_{max}=0.1\,h$/Mpc is consistent with the linear theory predictions
over all the redshifts considered to $1\sigma$. In fact, $\sigma_v$ in
the Taruya$^{++}$ and Kaiser models show a different redshift
dependence compared to the fiducial, although it should be exactly
degenerate with the linear growth factor $D(z)$ according to the
definition below Eq.~\ref{streaming}. This explains why the
Taruya$^{++}$ power spectrum produces a growth rate that is closest to
the fiducial cosmology when $\sigma_v$ is left as a free parameter;
the functional dependence is genuinely distinct from the linear theory
as we shall later see.

\begin{figure}
\begin{center}
\includegraphics[width=\columnwidth]{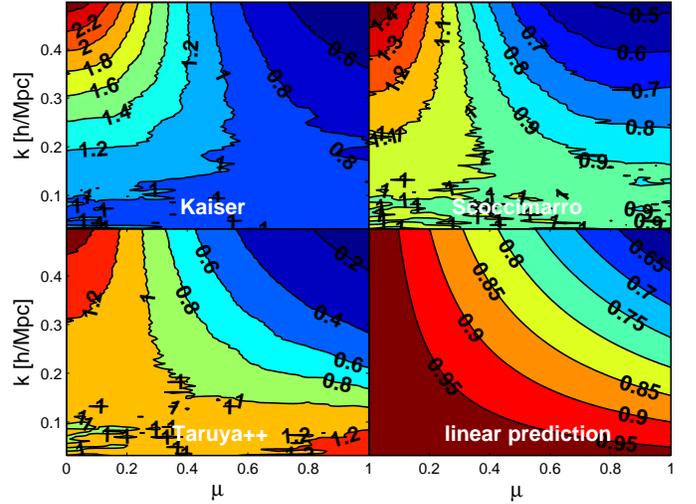}
\caption{2D isocontours of $F(k,\mu)$, using the Kaiser, Scoccimarro,
and Taruya$^{++}$ models of the redshift space power spectrum at $z=0$
using the exact N-body expressions for $P_{\delta\delta},
P_{\delta\theta}$ and $P_{\theta\theta}$.  The lower right corner
shows the prediction of linear theory, i.e.\ the standard exponential
form that is used in the streaming model.}

\label{fig:F_contours} 
\end{center}
\end{figure}

If the damping term is poorly described by an exponential, then what
is the correct functional form that must be applied to these RSD
models to reproduce the simulated redshift space power spectrum?
Defining the damping term by $F(k,\mu) =
P^s_{nbody}(k,\mu)/P^s_{model}(k,\mu)$, we divide the non-linear
redshift space power spectrum as measured in the simulations by the
models that we have considered so far. We caution that this function
can not be interpreted as the pairwise velocity PDF for the excellent
reasons outlined in~\cite{2004PhRvD..70h3007S}. For physical reasons, it
is likely that $F(k,\mu)$ is composed of a series of convolutions of
different functions at each scale and the true pairwise velocity PDF
is buried in this combination. 

Nonetheless it is instructive to characterise the functional form
required for these analytic models to succeed in describing the
redshift space power spectrum. Figure~\ref{fig:F_contours} shows
$F(k,\mu)$ for the Kaiser, Scoccimarro and Taruya$^{++}$ models. For
comparison, we have also shown the linear theory prediction for
$F(k,\mu)$ -- the standard exponential in $(fk\mu\sigma_v)^2$ -- in
the lower right corner of Fig.~\ref{fig:F_contours}.  A vague
similarity between these linear theory curves and those of the
Scoccimarro and Taruya$^{++}$ models only begins to appear for $\mu >
0.5$ and $k > 0.2 \,h$/Mpc. Furthermore, the damping is actually an
enhancement for much of the $k$-$\mu$ space, showing amplifications of
50-150\%.  As the RSD modelling improves, the maximum amplitudes of
$F(k,\mu)$ are smaller, with Scoccimarro and Taruya$^{++}$ models
having up to three times smaller deviations. This suggests that we can
build on their success by further modelling the non-linearities and
coupling between the density and velocity fields, here through an
empirical factor.

\begin{figure}
\begin{center}
\includegraphics[width=\columnwidth]{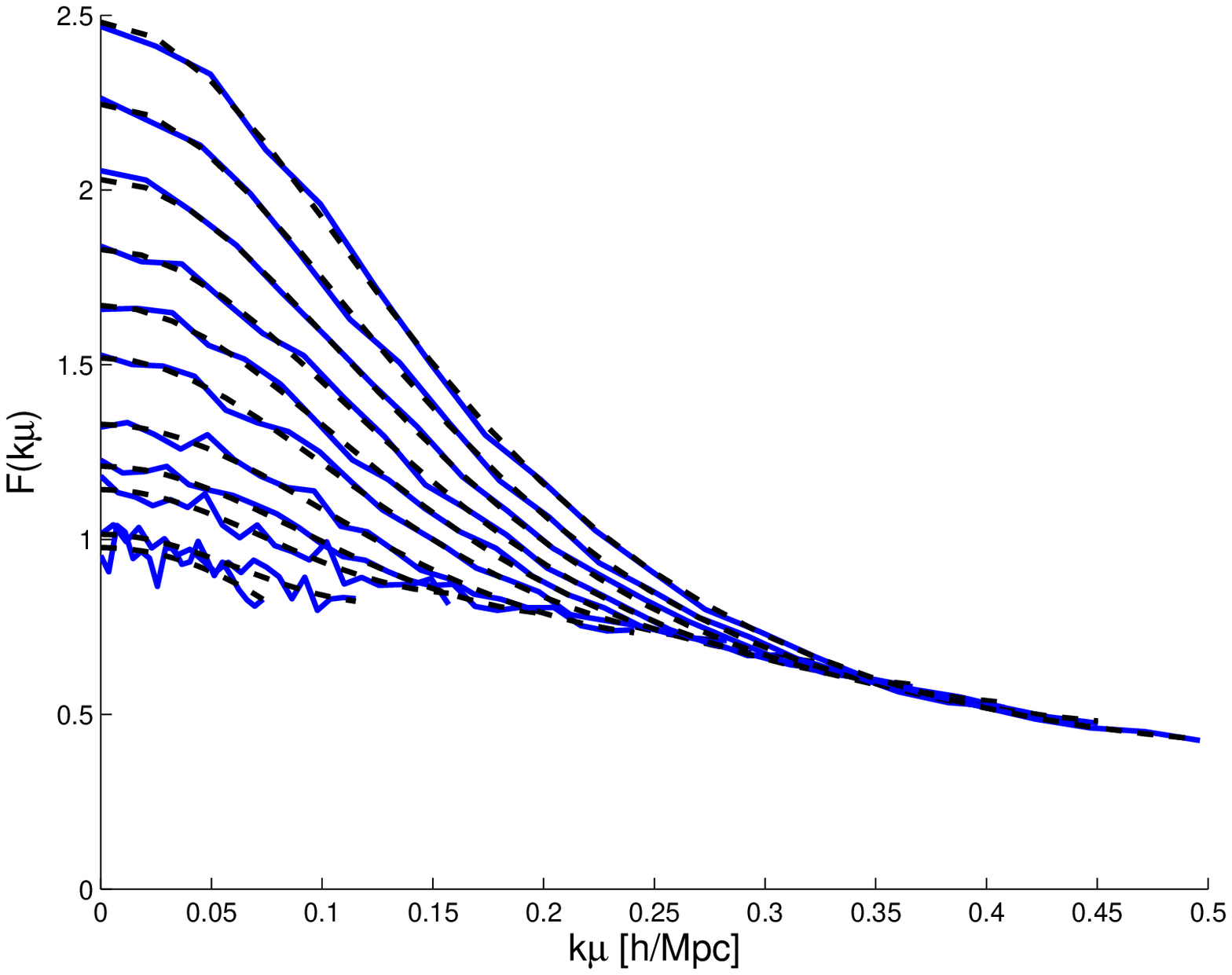}
\includegraphics[width=\columnwidth]{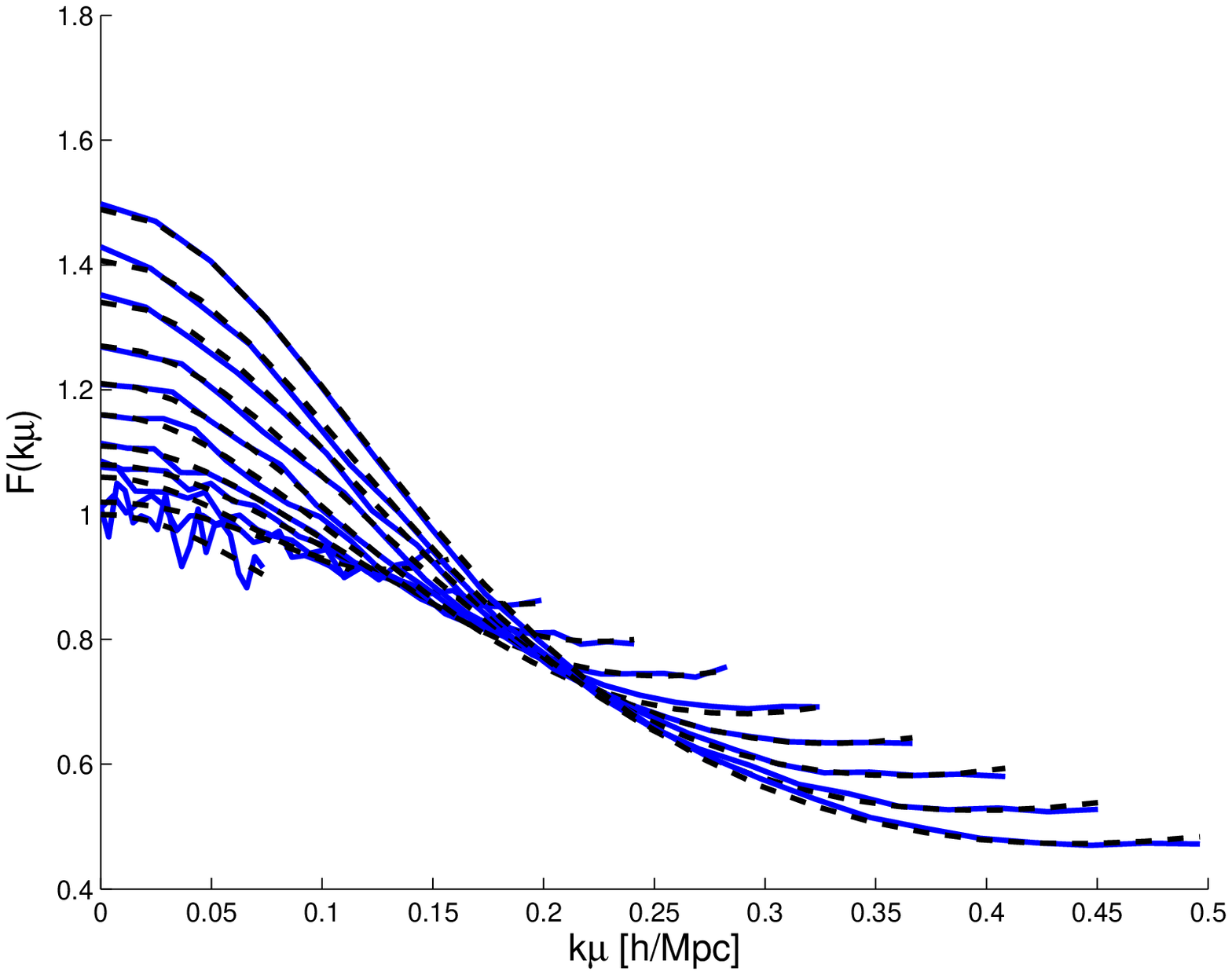}
\includegraphics[width=\columnwidth]{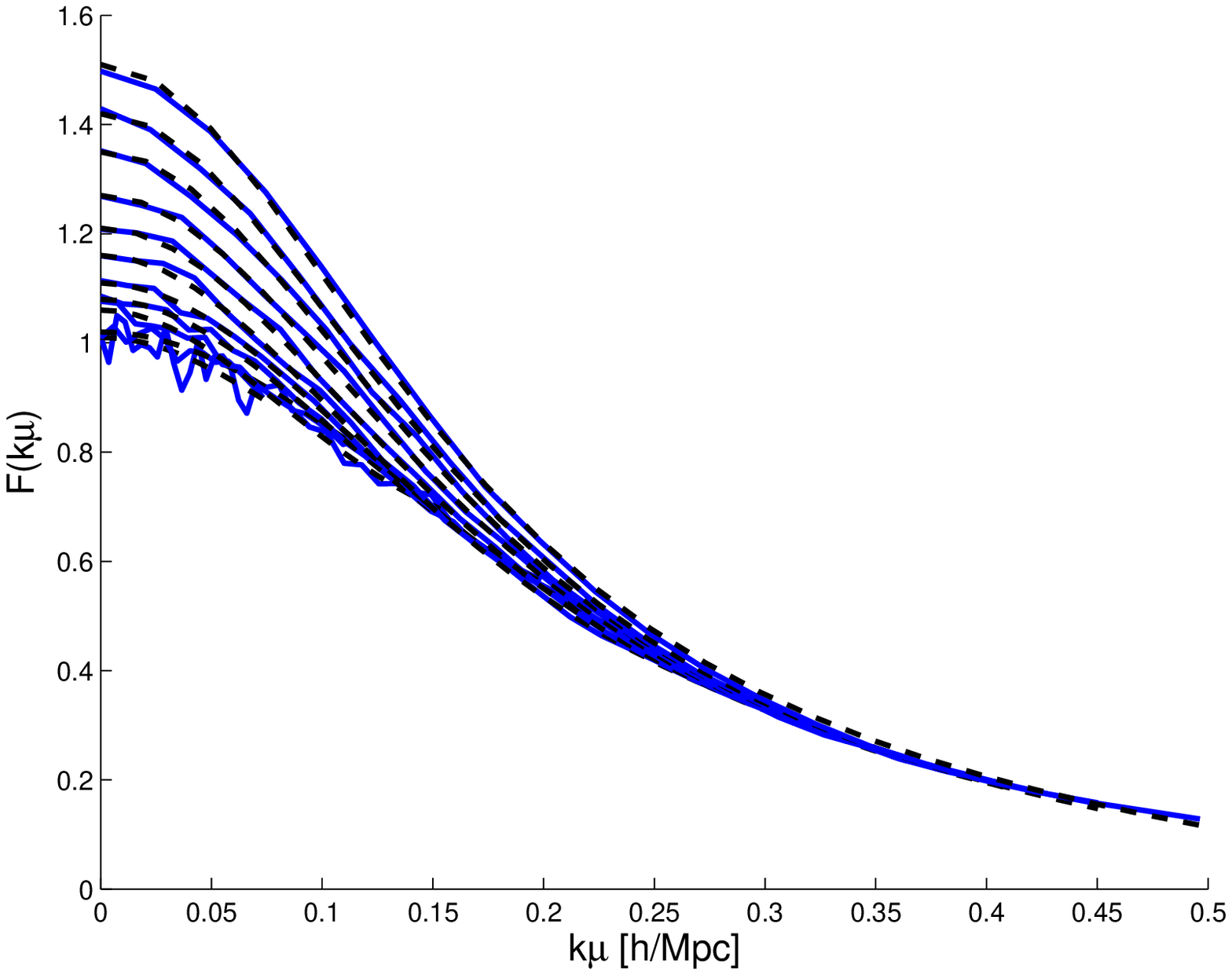}
\caption{Angular dependence factor $F(k,\mu)$ required for Kaiser
  (top), Scoccimarro (middle) and Taruya$^{++}$ (bottom) models to
  exactly describe the redshift space power spectrum at $z=0$. The
  blue curves are values of $F(k,\mu)$ as measured directly from
  simulations and the black dashed lines are the best fitting curves
  of the functional form in Eq.~\ref{Ffit} for a constant value of
  $k$, ranging from $0.07 < k < 0.5\,h$/Mpc, from bottom curve to top.}
\label{fig:fittingfunc}
\end{center}
\end{figure}

Since we are dealing with redshift distortions, $F(k,\mu)$ must enter as a 
function of $k\mu$, i.e.\ $k_z$.  We find that the curves of constant $k$ 
can be described out to $k\mu=0.5$ by 
\begin{equation}
F(k,\mu) = \frac{A}{1+Bk^2\mu^2}+Ck^2\mu^2, 
\label{Ffit}
\end{equation}
where A, B and C are free parameters for each of the models. In
general, the parameters are different for each model, but our
preliminary investigations suggest that there is some sort of
relationship between these values. Expansion of the denominator in
$k\mu$ would lead to $\mu^4$ (and higher order) terms, reminiscent of
the quartic terms in~\cite{1103.3614}. In contrast, we find that the
amplitude of $F(k,\mu)$ is non-negative on small scales but unity on
large scales for $\mu=0$, as required. The non-linear, FoG effects are
responsible for amplifying the RSD signal perpendicular to the line of
sight on small scales, and this is accurately captured by this
form. Also noteworthy is the necessity of {\em three\/} free
parameters -- the stretching and squashing effects cannot be
simultaneously described with a single parameter as in linear theory
damping. If we drop the extra term in C, we find that $F(k,\mu)$ falls
off too rapidly to describe regions of large $k\mu$ at small scales.
The fitting form agrees with the simulations at the percent level
(modulo the small $k$ oscillations discussed below). Although the
exact provenance of Eq.~\ref{Ffit} is the topic of future work, the
term in C arises because of extra contributions to the pairwise
velocity PDF beyond the usual exponential distibution. Futher work is
also required to test this functional form at other redshifts, for the
moment we have restricted ourselves to considering $z=0$.

Figure~\ref{fig:fittingfunc} shows that in general $F(k,\mu) = 1$ only
when both $k$ and $\mu$ are small. In fact, on extremely large scales
($k < 0.05\,h$/Mpc), $F(k,\mu) < 1$, but it was not possible to infer
the correct form of damping in this regime because the range of $k\mu$
considered is then necessarily small. Of greater concern, however, is
that these curves display some oscillatory behaviour that raises the
issue that incorrect modelling of RSD may contribute to some bias in
the baryon acoustic feature. We have confirmed that when $F(k,\mu)$ is
kept at a constant value of $\mu$ (Fig.~\ref{fig:fittingfunc} shows
curves of constant $k$) that the oscillations vary at a level above
the sample variance (the latter is no more than 2\% on these scales
and the oscillations are 6\% above the background). Moreover, the
shape of the oscillations vary with the RSD model indicating that a
genuine bias may be present. This will be explored further in future
work, requiring larger simulations. This will be explored further in
future work, requiring larger simulations.

\section{Testing gravity}\label{sec:extended} 

We now consider extending the fitting process to include the full set
of parameters that can be reasonably constrained by RSD alone, such
that $\{\gamma, b, \omega_m, w_0\}$ are now free parameters, but we
keep the variation in $w$ fixed at $w_a$ = 0. The parameter $\omega_m$
is defined in terms of the matter energy density such that $\omega_m =
\Omega_m h^2$. We only use the redshift bins at $z=0.5, 1$ as
appropriate for current and future surveys and limit ourselves to
using the least biased model of RSD out of each class of models that
we have considered. These are the streaming model with non-linear
matter power spectrum, the Scoccimarro formula with SPT power spectra
and the Taruya$^{++}$ model. With the first two, we use linear theory
damping; as the previous section showed, the constraints in the Kaiser
and Scoccimarro models are fairly insensitive to the value of
$\sigma_v$ used, but we have used an empirical damping term for the
Taruya$^{++}$ because this seems to minimise the bias in the derived
values for $f$. We allow the linear bias to vary to soften the
constraints as a best-case scenario, since we would expect some
covariance between the two in a real galaxy survey. We have also
checked that this did not negatively affect the ability of any of
these three RSD models to correctly model the growth rate. In all of
the models considered, the linear bias is quite tightly constrained
and is consistent with $b=1$ at all redshifts. Furthermore, it is the
shape of the redshift space power spectrum, rather than the amplitude,
that most informs us about cosmology.

\begin{figure*}
\begin{center}
\includegraphics[width=2\columnwidth]{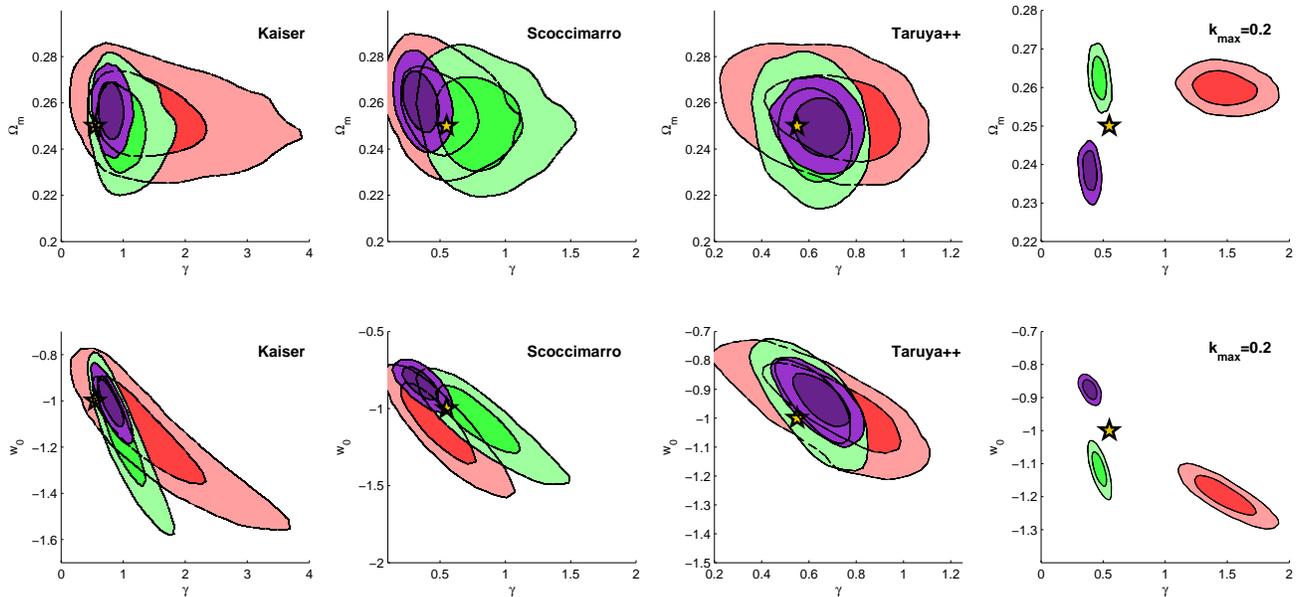}
\caption{1 and 2 $\sigma$ posterior distributions on $\gamma$ after
  extending the parameter set to include $\{\gamma, b, \omega_m,
  w_0\}$ for the streaming model (leftmost column), the Scoccimarro
  model (second column), and Taruya$^{++}$ model (third column) for
  $k < 0.1\,h$/Mpc. In each of these figures, the posteriors are
  coloured according to redshift, $z=0.5$ (light green), $z=1$ (red)
  and joint fits over both redshifts are in dark purple.  Note the
  change in scales between the figures.  The rightmost column shows
  the Kaiser (red), Scoccimarro (light green), and Taruya$^{++}$ (dark
  purple) models using $k < 0.2\,h$/Mpc at $z=1$. The fiducial
  parameters are marked with a star.  }
\label{fig:gamma_all}
\end{center}
\end{figure*}

Figure~\ref{fig:gamma_all} shows the 1 and 2$\sigma$ posterior
distributions in both the $\gamma-\om$ (top row) and $\gamma-w_0$
(bottom row) parameter spaces, using out to $k_{max}=0.1\,h$/Mpc.
Recall that $\gamma$ is the gravitational growth index and provides a
test of general relativity, which basically predicts $\gamma=0.55$.

Treating the power spectra measured at $z=0.5$ and $z=1$ individually,
and restricting to $k<0.1\,h$/Mpc, we found that the joint constraints
on $\gamma$ and $\om$ were mostly unbiased; the posterior
distributions of both the Scoccimarro ansatz and Taruya$^{++}$ model
contained the fiducial values within their $1\sigma$ regions, while
for the Kaiser formula the fiducial values were within the 1$\sigma$
region at $z=1$ and within the 2$\sigma$ limit at $z=0.5$. The
contours in the $\gamma-w_0$ plane show similar characteristics, with
slightly more bias.  The marginalised posterior distributions on
$\gamma$ for both the Scoccimarro and Taruya$^{++}$ models are
correct to within 1$\sigma$, but these bounds are now generous enough
to allow a wide swathe of cosmological models.

The constraints on $\gamma$ from RSD substantially weaken when using
high redshift slices, as expected from Eq.~\ref{eq:ferror} as
$\Omega_m(a)\to1$. The width of the $1\sigma$ contours on $\gamma$
also prevents us from drawing any interesting conclusions about
gravity with RSD alone, although in a simplistic sense, the effective
volume of the simulations is $\approx 67.5\;[$Gpc/$h]^3$ if we ignore
the contributions of larger scale modes.  Low redshift surveys capable
of accurately measuring $\gamma$, together with accurate modelling of
RSD, will be important if we wish to test gravity and cosmology.

One of the advantages of using the gravitational growth index is that
we expect to measure the same value of $\gamma$ at each redshift for
scale independent theories of gravity and we may combine observations
from several redshift bins to strengthen our constraints. We have
combined the power spectra in these redshift bins, neglecting any
covariance between them. This gives the purple contours seen in
Fig.~\ref{fig:gamma_all}. But none of the models have benefitted from
such treatment, despite the substantial improvement in the statistical
error. This is partly because it would be preferable to fit each
redshift slice individually since the sensitivity of the results to
$k_{max}$ is different at each redshift, but also indicates that some
of these models are becoming less valid as the redshift distribution
becomes more non-linear. This is particularly true for the Scoccimarro
ansatz, whose joint posteriors are the most inconsistent with the two
individual distributions.

Attempting to use power spectrum information out to
$k_{max}=0.2\,h$/Mpc with any of these models severely exacerbates the
bias, even as the statistical errors shrink.  For example, the
rightmost column of Fig.~\ref{fig:gamma_all} shows that at $z=1$ where
the redshift space density field might be expected to be reasonably
linear, the posterior distributions are far more than 2$\sigma$ away
from their fiducial values regardless of the model. However, the
Taruya$^{++}$ model could certainly benefit from using a different
estimate of $P_{\delta\delta},\, P_{\delta\theta}$ and
$P_{\theta\theta}$ on small scales.

\section{Conclusions} \label{sec:concl}

With an combined volume of 67.5 [Gpc/$h]^3$, we have tested the
robustness of the Kaiser limit, Scoccimarro ansatz, Lagrangian and
Standard perturbation theories and Taruya$^{++}$ models of redshift
space distortions, with a number of variations on the form of the
damping term. Of all the linear, quasi-linear and non-linear models
for the redshift space power spectrum that we have considered, we
found that the least biased of these for determining the growth rate
$f$ was the model proposed by~\cite{2010PhRvD..82f3522T} based on the
Taruya$^{++}$ model if we use standard perturbation theory to evaluate
the real space components and restrict the scales used to
$k<0.1\,h$/Mpc.  However, none of the models delivered accurate
results if we extend them to $k=0.2\,h$/Mpc.

Even with sophisticated perturbation theory schemes, all the models
that we have considered fail to describe correctly the angular
dependence of the redshift space power spectrum in the quasi-linear
regime and mischaracterise the non-linear Finger-of-God effects.
Instead, we present a fitting formula for the angular dependence
factor that, in comparison to simulations, accounts accurately for
both damping and FoG effects. Further work is needed to relate this to
underlying theory and to test its universality.  One issue of concern
is the possible interaction of the redshift space distortion modelling
with extraction of the true baryon acoustic oscillation scale.

The simulations provide the redshift space power spectrum of the dark
matter haloes.  As work such as \cite{2011ApJ...726....5O} shows, the
behavior of galaxies relative to dark matter can have added
complications. We cannot fully account for the difference between what
is measured in a galaxy survey and models of redshift space
distortions until we also consider the problems of biased tracers, as
well as wide angle redshift space distortions within perturbation
theory and additional astrophysical effects such as non-zero neutrino
masses. Other areas of interest include using redshift space
distortions to constrain the metric potentials, instead of the
gravitational growth index, to probe the gravitational model.

In terms of testing gravity from the growth rate, we found the results
are not robust with respect to inaccurate redshift space distortion
modeling.  Bias in the confidence regions for the gravitational growth
index $\gamma$ and other cosmological parameters can be appreciable,
making it dangerous to use modes as large as $k=0.1\,h$/Mpc, even for
a rather conservative set of cosmological parameters.  In fact, the set 
$\{f,b,\om, w_0\}$, is substantially smaller than those 
cosmological parameters that must be
considered~\citep{2010PhRvD..81d3512S} if we are to rule out general
relativity with any confidence.

This might be ameliorated by combination of RSD with other probes,
though there is also the danger that this will merely reduce the
statistical errors. If the bias in the RSD results is particularly
strong, as is the case with $k_{max} = 0.2 \, h$/Mpc, there is the
additional risk that the RSD dataset will simply be inconsistent with
and pull the results of other measurements or that the resultant distribution
will be multimodal. Including the parameters from our angular dependence
factor, or its future forms, may provide a robust approach to mapping
growth and gravity with redshift space distortions.

\section*{Acknowledgments}
JK thanks Berkeley Lab and the Berkeley Center for Cosmological
Physics, Los Alamos National Laboratory, the Institute of Astronomy,
Cambridge and Swinburne University of Technology for hospitality while
this work was conducted. JK would like to thank Chris Blake for
helpful comments on a draft of this article. This work has been
supported in part by the Director, Office of Science, Office of High
Energy Physics, of the U.S.\ Department of Energy under Contract
No.\ DE-AC02-05CH11231, and the World Class University grant
R32-2009-000-10130-0 through the National Research Foundation,
Ministry of Education, Science and Technology of Korea. JK and GFL
acknowledge support from ARC Discovery Project DP0665574, and
significant computational resources through the INTERSECT/NCI partner
share.

\end{document}